\documentclass[a4paper,12pt]{amsart}
\usepackage{amssymb}
\usepackage[colorlinks=true]{hyperref}
\usepackage{tikz}\usetikzlibrary{decorations.markings}
\allowdisplaybreaks[2]
\title[Sum rule for the eight-vertex model at $\eta=\pi/3$]{Sum rule for the eight-vertex model on its combinatorial line}
\newtheorem{conjecture}{Conjecture}

\newtheorem*{theorem*}{Theorem}
\newtheorem*{proposition*}{Proposition}
\newtheorem*{corollary*}{Corollary}
\long\def\rem#1{}
\newcommand\bra[1]{\left<#1\,\right|}
\newcommand\ket[1]{\left|\,#1\right>}
\newcommand{\braket}[2]{\left< #1 \vphantom{#2}\, \right|
\! \left. #2 \vphantom{#1} \right>}%
\def\eqbreak#1\\{\\*
\noalign{\vbox\bgroup\hfill$\def\eqbreak{$\egroup\nobreak\vbox\bgroup\hfill$}\displaystyle #1$\egroup}
}
\renewcommand\th{\vartheta}
\newcommand\ph{\varphi}
\newcommand\Pf{\mathop{\mathrm{Pf}}}
\newcommand\A{\mathrm{A}}
\newcommand\B{\mathrm{B}}
\newcommand\X{\mathrm{X}}
\newcommand\Z{\mathrm{Z}}
\newcommand\M{\mathrm{M}}
\renewcommand\S{\mathrm{S}}
\newcommand\f{\mathrm{f}}
\newcommand\g{\mathrm{g}}
\newcommand\h{\mathrm{h}}
\renewcommand\H{\mathrm{H}}
\author[P.~Zinn-Justin]{Paul Zinn-Justin}
\address{P.~Zinn-Justin, UPMC Univ Paris 6, CNRS UMR 7589, LPTHE,
75252 Paris Cedex, France}
\email{pzinn@lpthe.jussieu.fr}
\thanks{PZJ is supported in part by ERC grant 278124 ``LIC''.
PZJ would like to thank R.~Weston for his help 
in the framework of a parallel project, V.~Bazhanov and Mangazeev
for useful conversations, H.~Rosengren
for explaining his work \cite{rosengrenb} as well
as further unpublished work, and P.~Di Francesco for discussions.
Part of this work was performed during the author's stay at MSRI, Berkeley.}
\date{\today}
\begin{document}
\begin{abstract}
We investigate the conjectured ground state eigenvector
of the 8-vertex model inhomogeneous transfer matrix on its combinatorial
line, i.e., at $\eta=\pi/3$,
where it acquires a particularly simple form. We compute the
partition function of the model on an infinite cylinder with
certain restrictions on the inhomogeneities, and taking the homogeneous limit,
we obtain an expression for the squared norm of the ground state of the
XYZ spin chain as a solution of a differential recurrence relation.
\end{abstract}
\maketitle
\section{Introduction}
The purpose of this article is to investigate
the {\em inhomogeneous eight-vertex model}\/ on a particular one-dimensional
family of the globally defined parameters of the model, namely,
with the conventions of Baxter \cite{Baxter}, when $\eta=\pi/3$. 
More precisely, we study a certain eigenvector (conjecturally,
the ground state eigenvector in an appropriate range of parameters) 
of the transfer matrix of this 
model with periodic boundary conditions and an odd number of sites.
Ultimately,
the goal is to compare with some observations and conjectures 
\cite{BM-P6b,RS-XYZ}
made for the homogeneous eight-vertex model and the closely related
XYZ spin chain, but the introduction of inhomogeneities (spectral parameters)
turns out to be quite useful, as was previously found for the six-vertex
model \cite{artic31,artic36,artic42}, a special case of the
eight-vertex model.

In this section we briefly describe the model and some conjectured
properties at $\eta=\pi/3$. The rest of this paper is devoted
to showing how some of these properties arise from specializing formulae for
the inhomogeneous model. The main object of study will be the ``partition
function'' of the model on an infinite cylinder (equivalently,
a quadratic functional of the ground state eigenvector), for which we derive
an inhomogeneous sum rule (with a certain restriction on
the inhomogeneities, which we call ``half-specialization'') and a detailed
discussion of its homogeneous limit.
Note that this paper is not meant
to be fully mathematically rigorous; firstly, it is based on a conjecture
(Conj.~\ref{conj})
which we hope to prove in future work \cite{articxx}. Secondly,
some calculations involving theta and elliptic functions are skipped;
though they are in principle elementary, they can be quite tedious.

It should be noted that a special case of the eight-vertex model
on its combinatorial line, namely the six-vertex model at $\Delta=-1/2$,
is much better understood \cite{Strog-odd,BdGN-XXZ-ASM-PP,RS-spin,artic36,artic42}, 
and in this case many formulae of this work
are already known and proved; we provide in appendix~\ref{zetazero} the connection
to earlier work by taking the limit to the six-vertex point.

\subsection{Inhomogeneous eight-vertex transfer matrix}
The eight-vertex model is a two-dimensional statistical lattice model
defined on the square lattice by the assignment of arrows to each edge
of the lattice, according to eight possible local configurations
around a vertex:
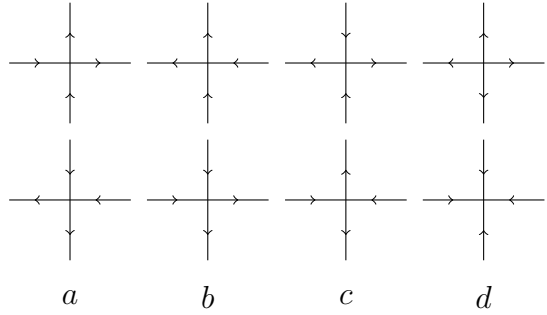
\begin{center}
\begin{tikzpicture}[x=0.8cm,y=0.8cm]
\matrix[column sep=2mm,row sep=2mm]{
\draw[decoration={markings,mark = at position 0.5 with { \arrow{>} }},postaction={decorate}]  (-1,0) -- (0,0); 
\draw[decoration={markings,mark = at position 0.5 with { \arrow{>} }},postaction={decorate}]  (0,0) -- (1,0); 
\draw[decoration={markings,mark = at position 0.5 with { \arrow{>} }},postaction={decorate}]  (0,-1) -- (0,0); 
\draw[decoration={markings,mark = at position 0.5 with { \arrow{>} }},postaction={decorate}]  (0,0) -- (0,1); 
&
\draw[decoration={markings,mark = at position 0.5 with { \arrow{<} }},postaction={decorate}]  (-1,0) -- (0,0); 
\draw[decoration={markings,mark = at position 0.5 with { \arrow{<} }},postaction={decorate}]  (0,0) -- (1,0); 
\draw[decoration={markings,mark = at position 0.5 with { \arrow{>} }},postaction={decorate}]  (0,-1) -- (0,0); 
\draw[decoration={markings,mark = at position 0.5 with { \arrow{>} }},postaction={decorate}]  (0,0) -- (0,1); 
&
\draw[decoration={markings,mark = at position 0.5 with { \arrow{<} }},postaction={decorate}]  (-1,0) -- (0,0); 
\draw[decoration={markings,mark = at position 0.5 with { \arrow{>} }},postaction={decorate}]  (0,0) -- (1,0); 
\draw[decoration={markings,mark = at position 0.5 with { \arrow{>} }},postaction={decorate}]  (0,-1) -- (0,0); 
\draw[decoration={markings,mark = at position 0.5 with { \arrow{<} }},postaction={decorate}]  (0,0) -- (0,1); 
&
\draw[decoration={markings,mark = at position 0.5 with { \arrow{<} }},postaction={decorate}]  (-1,0) -- (0,0); 
\draw[decoration={markings,mark = at position 0.5 with { \arrow{>} }},postaction={decorate}]  (0,0) -- (1,0); 
\draw[decoration={markings,mark = at position 0.5 with { \arrow{<} }},postaction={decorate}]  (0,-1) -- (0,0); 
\draw[decoration={markings,mark = at position 0.5 with { \arrow{>} }},postaction={decorate}]  (0,0) -- (0,1); 
\\
\draw[decoration={markings,mark = at position 0.5 with { \arrow{<} }},postaction={decorate}]  (-1,0) -- (0,0); 
\draw[decoration={markings,mark = at position 0.5 with { \arrow{<} }},postaction={decorate}]  (0,0) -- (1,0); 
\draw[decoration={markings,mark = at position 0.5 with { \arrow{<} }},postaction={decorate}]  (0,-1) -- (0,0); 
\draw[decoration={markings,mark = at position 0.5 with { \arrow{<} }},postaction={decorate}]  (0,0) -- (0,1); 
&
\draw[decoration={markings,mark = at position 0.5 with { \arrow{>} }},postaction={decorate}]  (-1,0) -- (0,0); 
\draw[decoration={markings,mark = at position 0.5 with { \arrow{>} }},postaction={decorate}]  (0,0) -- (1,0); 
\draw[decoration={markings,mark = at position 0.5 with { \arrow{<} }},postaction={decorate}]  (0,-1) -- (0,0); 
\draw[decoration={markings,mark = at position 0.5 with { \arrow{<} }},postaction={decorate}]  (0,0) -- (0,1); 
&
\draw[decoration={markings,mark = at position 0.5 with { \arrow{>} }},postaction={decorate}]  (-1,0) -- (0,0); 
\draw[decoration={markings,mark = at position 0.5 with { \arrow{<} }},postaction={decorate}]  (0,0) -- (1,0); 
\draw[decoration={markings,mark = at position 0.5 with { \arrow{<} }},postaction={decorate}]  (0,-1) -- (0,0); 
\draw[decoration={markings,mark = at position 0.5 with { \arrow{>} }},postaction={decorate}]  (0,0) -- (0,1); 
&
\draw[decoration={markings,mark = at position 0.5 with { \arrow{>} }},postaction={decorate}]  (-1,0) -- (0,0); 
\draw[decoration={markings,mark = at position 0.5 with { \arrow{<} }},postaction={decorate}]  (0,0) -- (1,0); 
\draw[decoration={markings,mark = at position 0.5 with { \arrow{>} }},postaction={decorate}]  (0,-1) -- (0,0); 
\draw[decoration={markings,mark = at position 0.5 with { \arrow{<} }},postaction={decorate}]  (0,0) -- (0,1); 
\\
\node {$a$};&
\node {$b$};&
\node {$c$};&
\node {$d$};
\\
};
\end{tikzpicture}
\end{center}
They are given Boltzmann weights denoted by $a,b,c,d$
which are parameterized as follows:
\begin{equation}\label{boltz}
\begin{split}
a(x)&=\th_4(2\eta,p^2) \th_4(x,p^2)\th_1(x+2\eta,p^2)\\
b(x)&=\th_4(2\eta,p^2) \th_1(x,p^2)\th_4(x+2\eta,p^2)\\
c(x)&=\th_1(2\eta,p^2) \th_4(x,p^2)\th_4(x+2\eta,p^2)\\
d(x)&=\th_1(2\eta,p^2) \th_1(x,p^2)\th_1(x+2\eta,p^2)
\end{split}
\end{equation}
where $x$ is the spectral parameter and $p=e^{\mathrm{i} \pi \tau}$,
$\mathrm{Im}\ \tau>0$, is the elliptic nome.
The weights have period $2\pi$ and pseudo-period $2\pi\tau$, i.e.,
they are multiplied by a {\em common}\/ factor when $x$ is replaced
with $x+2\pi\tau$.

Ordering the edge states as $(\uparrow,\downarrow)$ and $(\rightarrow,\leftarrow)$, these weights
can be encoded into the $R$-matrix
\[
R(x)=
\begin{pmatrix}
a(x)&0&0&d(x)\\
0&b(x)&c(x)&0\\
0&c(x)&b(x)&0\\
d(x)&0&0&a(x)
\end{pmatrix}
\]
We shall also need in what follows the $\check R$-matrix defined as
\[
\check R(x)=\mathcal{P} R(x)=
\begin{pmatrix}
a(x)&0&0&d(x)\\
0&c(x)&b(x)&0\\
0&b(x)&c(x)&0\\
d(x)&0&0&a(x)
\end{pmatrix}
\]
where $\mathcal{P}$ permutes factors of the tensor product.
\rem{what are the linear relations between $a,b,c,d$?}

The Boltzmann weights satisfy the Yang--Baxter equation and unitarity
equation;
in terms of $\check R$, these are expressed as 
\begin{equation}\label{ybe}
\check R_{i,i+1}(x)\check R_{i+1,i+2}(x+y) \check R_{i,i+1}(y)
=
\check R_{i+1,i+2}(y)\check R_{i,i+1}(x+y) \check R_{i+1,i+2}(x)
\end{equation}
\rem{should also provide picture since we need another version of it, really}
and
\begin{equation}\label{unit}
\check R(x)\check R(-x)=r(x)r(-x)
\,1
\end{equation}
where $r(x)=\th_4(0;p^2)\th_1(x-2\eta;p^2)\th_4(x-2\eta;p^2)$.

We consider here the model in size $L$ with periodic boundary conditions in the horizontal direction, i.e.,
with the geometry of a cylinder of width $L$. The state of the $L$ vertical edges at same height on the
cylinder are encoded by a sequence in $\{\uparrow,\downarrow\}^L$.
The {\em transfer matrix}\/ is a $2^L\times 2^L$ matrix, or equivalently
an operator on $(\mathbb{C}^2)^{\otimes L}$
with its standard basis indexed by $\{\uparrow,\downarrow\}^L$,
describing the transition from one row of vertical edges to the next; 
the fully inhomogeneous transfer matrix has the formal expression
\[
T_L(u|x_1,\ldots,x_L) = \mathrm{Tr}_0\ R_{01}(x_1-u)R_{02}(x_2-u)\ldots R_{0L}(x_L-u)
\]
where we use the following convention: the indices of operators $R$ (and all other
local operators) are the spaces on which they act in the tensor product $(\mathbb{C}^2)^{\otimes L}$.
$u,x_1,\ldots,x_L$ are spectral parameters of the model.
The system has rotational invariance in the sense that shifting cyclically
sites in the tensor product {\em and}\/ spectral parameters leaves $T_L$ invariant.
In what follows, all indices in $\{1,\ldots,L\}$
must be understood modulo $L$.

Finally we need Pauli matrices $\sigma^{x,y,z}$, which are local operators acting on one site;
we give alternate names to two of them.
The flip operator ($\sigma^x$ Pauli matrix)
is $F=\left(\begin{smallmatrix}0&1\\1&0\end{smallmatrix}\right)$
and the spin operator ($\sigma^z$ Pauli matrix)
is $\sigma=\left(\begin{smallmatrix}1&0\\0&-1\end{smallmatrix}\right)$.
Finally $\sigma^y=\left(\begin{smallmatrix}0&\mathrm{i}\\-\mathrm{i}&0\end{smallmatrix}\right)$.
Denote $F_{\ast}=\prod_{i=1}^L F_i$.

The transfer matrix $T_L$ is invariant by reversal of all spins, i.e.,
$[T_L,F_\ast]=0$.

\subsection{Combinatorial line}
{\em In all the rest of this paper, we assume that $L$ is an odd number, $L=2n+1$,
and that $\eta=\pi/3$}. This second condition is what we call ``combinatorial line'', because
of the occurrence of integer numbers in the ground state, as we shall see below.
The value $\eta=\pi/3$ was first noticed to have special significance by Baxter
\cite{Bax-review}; the importance of odd $L$ was emphasized by Stroganov \cite{Strog-8v}. More recently,
Razumov and Stroganov \cite{RS-XYZ} and Bazhanov and Mangazeev
\cite{BM-8v,BM-P6a,BM-P6b} studied the model with such conditions.
It is also known \cite{BM-8v,FH-susy2} that $\eta=\pi/3$ corresponds to
a supersymmetric point for the XYZ spin chain.

Although the work \cite{RS-XYZ} is mostly concerned with the homogeneous limit
(see below), the following conjecture is made there (translated into our present conventions):
the transfer matrix $T_L(u|x_1,\ldots,x_L)$ possesses the eigenvalue
\[
t_L(u|x_1,\ldots,x_L)=\prod_{i=1}^L (a(x_i-u)+b(x_i-u))
\]
In fact, this eigenvalue is found to be doubly degenerate; in \cite{RS-XYZ,BM-P6a,BM-P6b}, this degeneracy
is lifted by fixing the parity of the number of $\uparrow$ in the eigenvector. Here we find it more
convenient to
choose a different convention, which is to diagonalize simultaneously $F_\ast$.

Note the identities
at $\eta=\pi/3$:
\[
r(x)=\th_4(0,p^2)\th_1(x+\eta,p^2)\th_4(x+\eta,p^2)=a(x)+b(x)
\]

\subsection{Homogeneous limit}\label{sec:homlim}
If we assume that all $x_i$ are equal (homogeneous situation), then the transfer matrix $T_L$
commutes with the {\em XYZ Hamiltonian}, which can be written as
\[
H_L =-\frac{1}{2} \sum_{i=1}^L (J_4 \sigma_i^x \sigma_{i+1}^x + J_3 \sigma_i^y \sigma_{i+1}^y + J_2 \sigma_i^z \sigma_{i+1}^z )
\]
The numbering of the coupling constants will be explained later. The value $\eta=\pi/3$
implies that up to normalization, the three coupling constants can be expressed
in terms of a single quantity, which we choose to be
\[
\zeta=\left(\frac{\th_1(\eta;p^2)}{\th_4(\eta;p^2)}\right)^2
\]
If we choose $\tau$ purely imaginary, 
then as $p$ goes from $0$ to $1$, $\zeta$ goes
from $0$ to $1$.

\rem{should there be a sequence somewhere on various integer sequences? cf uniformize2d}

The coupling constants are given up to overall normalization by
\[
J_2=-\frac{1}{2}
\qquad
J_3=\frac{1}{1+\zeta}
\qquad
J_4=\frac{1}{1-\zeta}
\]
The XXZ Hamiltonian (corresponding to the six-vertex transfer matrix), is the 
case $\zeta=0$ (or $p=0$).
This case was already
studied in detail, as mentioned in the introduction.

Another special case is $\zeta\to 1$ (or $p\to1$), for which after rescaling the weights, $J_2=J_3=0$, so the
model becomes the Ising model, but with a $\sigma^x \sigma^x$ interaction.
The ground state becomes of course trivial; some details are provided in appendix~\ref{trivtrigo}.

The simple eigenvalue of the eight-vertex transfer matrix translates into a simple
eigenvalue of the Hamiltonian $H_L$, namely
\[
E_L = -\frac{L}{2} (J_2+J_3+J_4)
\]
It is conjectured to be the ground state eigenvalue of $H_L$.

Many remarkable observations were made on the corresonding eigenvector, $\Psi_L$ in \cite{RS-XYZ,BM-P6b}.
Its entries can be chosen to be {\em polynomials}\/ in $\zeta$, and the form of some of these polynomials
was conjectured. We shall not discuss these conjectures here. The values 
at $\zeta=0$ (XXZ model) of these
polynomials were calculated in \cite{artic42}.

In both \cite{RS-XYZ,BM-P6b}, the {\em squared norm}\/ of $\Psi_L$ was introduced:
\begin{equation}\label{sqnorm}
|\Psi_L|^2=\sum_{\alpha\in\{\uparrow,\downarrow\}^L} \Psi_{L;\alpha}^2
\end{equation}
where the normalization of the components is chosen so that they are coprime polynomials in $\zeta$,
and $\Psi_{L;\underbrace{\uparrow\cdots\uparrow}_n\underbrace{\downarrow\cdots\downarrow}_{n+1}}(\zeta=0)=1$.
An expression for $|\Psi_L|^2$ was conjectured 
in \cite{BM-P6b} in terms of certain polynomials, themselves
defined by differential recurrence relations which are special cases of certain B\"acklund transformations
for Painlev\'e VI. 
Since the formulae are rather complicated, we shall not write them out here
and derive our own (similar) formulae
by specializing inhomogeneous expressions. 

The main result of this paper is the factorization of this squared norm
into four factors, as summarized in sect.~\ref{summary}, which are
all determined by differential bilinear recurrence relations which are
given explicitly in appendix~\ref{moremesses}.

\section{Properties of the ground state eigenvector}\label{pties}
We consider once again the eigenvector equation in size $L=2n+1$
\begin{subequations}\label{eigall}
\begin{align}\label{eigenv}
T_L(u|x_1,\ldots,x_L)
\Psi_L(x_1,\ldots,x_L)
&=
t_L(u|x_1,\ldots,x_L)
\Psi_L(x_1,\ldots,x_L)
\\
F_\ast\Psi_L(x_1,\ldots,x_L)&=(-1)^n \Psi_L(x_1,\ldots,x_L)\label{eigflip}
\end{align}
\end{subequations}
for the inhomogeneous eight-vertex transfer matrix,
where we recall that $t_L(u|x_1,\ldots,x_L)=\prod_{i=1}^L r(x_i-u)$,
$r(x)=a(x)+b(x)=\th_4(0,p^2)\th_1(x+\eta,p^2)\th_4(x+\eta,p^2)$.
The choice of eigenvalue of $F_\ast$ will turn out convenient in what follows.

\subsection{Pseudo-periodicity}
Based on extensive study of the ground state entries
by computer for small sizes $L=3,5,7$, the following conjecture
seems valid:

\begin{conjecture}\label{conj}
The eigenvector equations \eqref{eigall} possess a solution
$\Psi_L(x_1,\ldots,x_L)$ whose entries are theta functions of degree $L-1=2n$
and nome $p^2$ in each variable $x_i$ (generically non zero and without common factor); 
i.e., they are holomorphic functions with pseudo-periodicity
property: 
\begin{subequations}\label{pseudoconj}
\begin{align}\label{pseudoconja}
\Psi_L(\ldots,x_i+2\pi \tau,\ldots)&=
p^{-4n} z_i^{2n}\prod_{j(\ne i)}z_j^{-1} \Psi_L(\ldots,x_i,\ldots)
\\\label{pseudoconjb}
\Psi_L(\ldots,x_i+\pi,\ldots)&=\prod_{j(\ne i)}\sigma_j\,
\Psi_L(\ldots,x_i,\ldots)
\end{align}
\end{subequations}
where the $\ldots$ mean unspecified variables $x_1$, $x_2$, etc,
$p=e^{\mathrm{i} \pi \tau}$ 
and $z_i=e^{-2\mathrm{i}x_i}$, $i=1,\ldots,L$.
\end{conjecture}

(Note that the factor $\prod_{j(\ne i)} z_j^{-1}$ is to be expected
since $\Psi_L$ only depends on differences of spectral parameters.
The factor $p^{-4n}$ can be absorbed in a redefinition of $\Psi_L$, but is
convenient. The factor $\prod_{j(\ne i)}\sigma_j$ is again expected from the properties
of the $R$-matrix by shift of $\pi$;
it could be absorbed in a simultaneous
redefinition of the $R$-matrix and of $\Psi_L$.)

Similar properties have been observed and (in some cases) proved for
models based on trigonometric or rational solutions of the Yang--Baxter
equation at special points of their parameter space \cite{artic31,artic32,PDF-open,PDF-open2,artic37,artic40,artic42,Cantini-open,dGPS}, except
the entries are ordinary polynomials of prescribed degree
(the main difficulty being to prove this degree).
In particular, in the limit $\zeta\to0$, $\Psi_L$ reduces to the
eigenvector of the inhomogeneous six-vertex transfer matrix
whose existence and uniqueness was
proved rigorously in \cite{artic42} and references therein. 
Therefore, if such a solution of \eqref{eigall} exists, it is necessarily
unique (for generic $p$) up to normalization. In principle this
normalization might contain a non-trivial function of the $x_i$,
which is why we added to the conjecture the fact that the entries
have no common factor.
So there remains only an arbitrary constant in
the normalization of $\Psi_L$, which will be fixed later.

\subsection{Exchange relation}
As a direct application of the Yang--Baxter equation,
we have the following intertwining relation:
\begin{equation}\label{intertw}
T_L(u|\ldots,x_{i+1},x_i,\ldots)
\check R_{i,i+1}(x_{i+1}-x_i)
=
\check R_{i,i+1}(x_{i+1}-x_i)
T_L(u|\ldots,x_i,x_{i+1},\ldots)
\end{equation}
(see Lemma 1 of \cite{artic31} for the same formula in a similar setting,
and its graphical proof).

Now apply $\Psi_L(x_1,\ldots,x_L)$ to Eq.~\eqref{intertw}
and use the eigenvalue equation~\eqref{eigenv}:
\begin{multline*}
T_L(u|x_1,\ldots,x_{i+1},x_i,\ldots,x_L)
\check R_{i,i+1}(x_{i+1}-x_i)\Psi_L(\ldots,x_i,x_{i+1},\ldots)
\\*
=
t_L(u|x_1,\ldots,x_L)
\check R_{i,i+1}(x_{i+1}-x_i)\Psi_L(\ldots,x_i,x_{i+1},\ldots)
\end{multline*}
$t_L(u|x_1,\ldots,x_L)$
being invariant by permutation of $x_i,x_{i+1}$, and $F_\ast$ commuting with $\check R_{i,i+1}$,
we conclude by the uniqueness of the solution of \eqref{eigall} that
\[
\check R_{i,i+1}(x_{i+1}-x_i)
\Psi_L(\ldots,x_i,x_{i+1},\ldots)
=
r_i(x_1,\ldots,x_L)
\Psi_L(\ldots,x_{i+1},x_{i},\ldots)
\]
where $r_i$ is some scalar function which is a ratio
of theta functions, but cannot have a non-trivial denominator
because it would be a common factor of the $\Psi_\alpha$, which would
contradict Conj.~\ref{conj}; so it is a
theta function
of degree 2 in $x_i,x_{i+1}$ (and zero in all others, hence a constant)
with given pseudo-periodicity property;
by applying the identity twice and using unitarity equation \eqref{unit},
we find $r_i(x_i,x_{i+1})r_i(x_{i+1},x_i)=r(x_i-x_{i+1})r(x_{i+1}-x_i)$.
The only theta function which divides the right hand side and has
the same pseudo-periodicity properties as $r_i(x_i,x_{i+1})$ is
$r(x_{i+1}-x_i)$; so $r_i(x_i,x_{i+1})=\pm r(x_{i+1}-x_i)$.
The simplest way to fix the sign is to use the $\zeta\to0$ limit
where it is known
\cite{artic42} that the correct sign is $+$. By continuity in $\zeta$, we have
in the end $r_i(x_i,x_{i+1})=r(x_{i+1}-x_i)$, so that
\begin{equation}\label{exch}
\check R_{i,i+1}(x_{i+1}-x_i)
\Psi_L(\ldots,x_i,x_{i+1},\ldots)
=
r(x_{i+1}-x_i)
\Psi_L(\ldots,x_{i+1},x_{i},\ldots)
\end{equation}

\subsection{Spin flip}
Next, note that weights $a(x)$ and $b(x)$ (resp.\ $c(x)$ and $d(x)$)
are exchanged by shift of $x$ by $\pi \tau$.
More precisely, we have the following identity:
\[
R_{01}(x+\pi \tau)=
-p^{-1} z
\,
F_1 
R_{01}(x)
F_1
\]
where $z=e^{-2\mathrm{i}x}$ and $F_1$ is the operator that flips
the second spin (of course the same would be true with $F_1$ replaced with
$F_0$, since the $R$ matrix commutes with $F_0F_1$).

Applying this to the transfer matrix, we find:
\begin{equation}\label{interflip}
T_L(\ldots,x_i+\pi\tau,\ldots)F_i=-p^{-1}z_i
\,F_i T_L(\ldots,x_i,\ldots)
\end{equation}
where we recall that $F_i$ flips spin $i$.

As in the previous section, apply $\Psi_L(x_1,\ldots,x_L)$ to 
Eq.~\eqref{interflip}
and use the eigenvalue equation \eqref{eigenv}:
\[
T_L(\ldots,x_i+\pi\tau,\ldots)
F_i
\Psi_L(\ldots,x_i,\ldots)=
-p^{-1}z_i\,
t_L(u|\ldots,x_i,\ldots)
F_i
\Psi_L(\ldots,x_i,\ldots)
\]
We have $t_L(u|\ldots,x_i+\pi\tau,\ldots)=-p^{-1}z_i t_L(u|\ldots,x_i,\ldots)$,
as should be, and $F_\ast$ and $F_i$ commute, so we conclude as before that 
\begin{equation}\label{preflip}
F_i \Psi_L(\ldots,x_i,\ldots)
= f_i(x_1,\ldots,x_L)
\Psi_L(\ldots,x_i+\pi\tau,\ldots)
\end{equation}
where $f_i(x_1,\ldots,x_L)$ is a scalar function
with the following properties: it is a ratio of theta functions,
but cannot have a non-trivial denominator
because it would be a common factor of the $\Psi_\alpha$, which would
contradict Conj.~\ref{conj}; so it is a holomorphic function,
with pseudo-periodicity properties determined
by shifting one of the $x_j$ by $\pi,\pi\tau$ in Eq.~\eqref{preflip}
and comparing with \eqref{pseudoconj};
we find
\begin{align*}
f_i(\ldots,x_i+\pi,\ldots)&=f_i(\ldots,x_i,\ldots)\\
f_i(\ldots,x_j+\pi,\ldots)&=-f_i(\ldots,x_j,\ldots)& j&\ne i\\
f_i(\ldots,x_i+2\pi\tau,\ldots)&=p^{4n} f_i(\ldots,x_i,\ldots)
\\
f_i(\ldots,x_j+2\pi\tau,\ldots)&=p^{-2} f_i(\ldots,x_j,\ldots)
& j&\ne i
\end{align*}
This fixes it to be 
$f_i(x_1,\ldots,x_L)=c\,
e^{2n\mathrm{i}x_i-\mathrm{i}\sum_{j(\ne i)} x_j}
$. By rotational invariance, the constant $c$ is independent of $i$.
Iterating Eq.~\eqref{preflip}
results in $f_i(\ldots,x_i,\ldots)f_i(\ldots,x_i+\pi\tau,\ldots)
=p^{4n}
e^{4n\mathrm{i}x_i-2\mathrm{i}\sum_{j(\ne i)} x_j}$,
which imposes that $c=\pm p^{n}$.
In order to fix the sign, we use the invariance by shift
of all the spectral parameters and the fact that $F_\ast\Psi_L=(-1)^n\Psi_L$
with $F_\ast=\prod_{i=1}^L F_i$ to conclude that
$c^L p^{-L(L-1)/2}=(-1)^n$ and therefore $c=(-p)^{n}$.

We finally obtain:
\begin{equation}\label{flip}
F_i \Psi_L(\ldots,x_i,\ldots)
= (-p)^{n}e^{2n\mathrm{i}x_i-\mathrm{i}\sum_{j(\ne i)} x_j}
\,\Psi_L(\ldots,x_i+\pi\tau,\ldots)
\end{equation}

\subsection{Wheel condition and recurrence relations}
\renewcommand\Im{\mathop{\mathrm{Im}}\nolimits}
We are now interested in the situation where two successive
spectral parameters have difference $2\eta$. In this paragraph,
we denote to simplify $T_L^+=T_L(u|\ldots,x,x+2\eta,\ldots)$
and $T_L^-=T_L(u|\ldots,x+2\eta,x,\ldots)$
where the two specialized spectral parameters are at sites $i,i+1$.
Applying the intertwining relation
\eqref{intertw} with $x_i=x+2\eta$, $x_{i+1}=x$,
we find:
\[
T_L^+ \check R_{i,i+1}(-2\eta)=\check R_{i,i+1}(-2\eta) T_L^-
\]
A direct calculation shows that
$\check R(-2\eta)=2\th_4(2\eta,p^2)\th_1(2\eta,p^2)
\th_4(0,p^2)P$ where $P$ is the projector
$P=\frac{1}{2}(1-\mathcal{P})=\frac{1}{2}\left(\begin{smallmatrix}
0&0&0&0\\
0&1&-1&0\\
0&-1&1&0\\
0&0&0&0
\end{smallmatrix}\right)$. Therefore the equality above
says that $T_L^+$ leaves $\Im P_{i,i+1}$ stable (and that
restricted to that subspace it is equal to the projection of $T_L^-$:
$T_L^+|_{\Im P_{i,i+1}}=P_{i,i+1}T_L^-|_{\Im P_{i,i+1}}$).

We shall need to check this explicitly.
Set $s=\ket{\uparrow\downarrow}-\ket{\downarrow\uparrow}=
\left(\begin{smallmatrix}0\\1\\-1\\0
\end{smallmatrix}\right)$ to be a generator of the image of projector $P$,
and compute $R_{0,i}(x)R_{0,i+1}(x+2\eta) s_{i,i+1}$, e.g.,
\tikzset{myarrow/.style={decoration={markings,mark = at position 0.5 with { \arrow[thick]{>} }},postaction={decorate}}}
\begin{align*}
\bra{\rightarrow}
R_{0,i}(x)R_{0,i+1}(x+2\eta)s_{i,i+1}
\ket{\leftarrow}
&=
\begin{tikzpicture}[baseline=0,draw/.append style=myarrow]
\draw (0,0) -- (1,0); \draw (3,0) -- (2,0);
\draw (1,-1) -- (1,0); \draw (2,0) -- (2,-1);
\draw (1,1) -- (1,0); \draw (2,0) -- (1,0); \draw (2,1) -- (2,0);
\end{tikzpicture}
+
\begin{tikzpicture}[baseline=0,draw/.append style=myarrow]
\draw (0,0) -- (1,0); \draw (3,0) -- (2,0);
\draw (1,-1) -- (1,0); \draw (2,0) -- (2,-1);
\draw (1,0) -- (1,1); \draw (1,0) -- (2,0); \draw (2,0) -- (2,1);
\end{tikzpicture}
\\
&\ -
\begin{tikzpicture}[baseline=0,draw/.append style=myarrow]
\draw (0,0) -- (1,0); \draw (3,0) -- (2,0);
\draw (1,0) -- (1,-1); \draw (2,-1) -- (2,0);
\draw (1,1) -- (1,0); \draw (1,0) -- (2,0); \draw (2,1) -- (2,0);
\end{tikzpicture}
-
\begin{tikzpicture}[baseline=0,draw/.append style=myarrow]
\draw (0,0) -- (1,0); \draw (3,0) -- (2,0);
\draw (1,0) -- (1,-1); \draw (2,-1) -- (2,0);
\draw (1,0) -- (1,1); \draw (2,0) -- (1,0); \draw (2,0) -- (2,1);
\end{tikzpicture}
\\
&=
(d(x)a(x+2\eta)-b(x)d(x+2\eta))
\ket{\downarrow\downarrow}
\\
&\ +
(a(x)c(x+2\eta)-c(x)b(x+2\eta))
\ket{\uparrow\uparrow}
\\
&=
0
\\
\bra{\rightarrow}
R_{0,i}(x)R_{0,i+1}(x+2\eta)s_{i,i+1}
\ket{\rightarrow}
&=
\begin{tikzpicture}[baseline=0,draw/.append style=myarrow]
\draw (0,0) -- (1,0); \draw (2,0) -- (3,0);
\draw (1,-1) -- (1,0); \draw (2,0) -- (2,-1);
\draw (1,0) -- (1,1); \draw (1,0) -- (2,0); \draw (2,1) -- (2,0);
\end{tikzpicture}
+
\begin{tikzpicture}[baseline=0,draw/.append style=myarrow]
\draw (0,0) -- (1,0); \draw (2,0) -- (3,0);
\draw (1,-1) -- (1,0); \draw (2,0) -- (2,-1);
\draw (1,1) -- (1,0); \draw (2,0) -- (1,0); \draw (2,0) -- (2,1);
\end{tikzpicture}
\\
&\ -
\begin{tikzpicture}[baseline=0,draw/.append style=myarrow]
\draw (0,0) -- (1,0); \draw (2,0) -- (3,0);
\draw (1,0) -- (1,-1); \draw (2,-1) -- (2,0);
\draw (1,0) -- (1,1); \draw (2,0) -- (1,0); \draw (2,1) -- (2,0);
\end{tikzpicture}
-
\begin{tikzpicture}[baseline=0,draw/.append style=myarrow]
\draw (0,0) -- (1,0); \draw (2,0) -- (3,0);
\draw (1,0) -- (1,-1); \draw (2,-1) -- (2,0);
\draw (1,1) -- (1,0); \draw (1,0) -- (2,0); \draw (2,0) -- (2,1);
\end{tikzpicture}
\\
&=
(a(x)b(x+2\eta)-c(x)c(x+2\eta))
\ket{\uparrow\downarrow}
\\
&\ +
(d(x)d(x+2\eta)-b(x)a(x+2\eta))
\ket{\downarrow\uparrow}
\\
&=
r(x+6\eta)r(x+2\eta)s_{i,i+1}
\end{align*}
and similarly with all arrows reversed.
Thus, $R_{0,i}(x)R_{0,i+1}(x+2\eta) s_{i,i+1}
=r(x+6\eta)r(x+2\eta)s_{i,i+1}\otimes 1_0$, and therefore, after shift
of $x\to x-u$, and use of $\eta=\pi/3$ to get rid of the $6\eta$, we find
\begin{equation}\label{recT}
T_L^+|_{\Im P_{i,i+1}}=r(x-u)r(x+2\eta-u) T_{L-2}
\end{equation}
where it is understood that $T_{L-2}$ acts only on sites distinct
from $i,i+1$.

Now apply 
$\Psi_{L-2}$ (with parameters $x_j$ except $x_i,x_{i+1}$)
tensor $s_{i,i+1}$
and use eigenvector equation \eqref{eigenv}:
\[
T_L^+ \Psi_{L-2}(\ldots)\otimes s_{i,i+1}
=r(x-u)r(x+2\eta-u) t_{L-2}(u|\ldots)
\Psi_{L-2}(\ldots)\otimes s_{i,i+1}
\]
By definition, $t_L(u|\ldots,x,x+2\eta,\ldots)=r(x-u)r(x+2\eta-u) t_{L-2}(u|\ldots)$.
Also, from Eq.~\eqref{eigflip},
$F_\ast \Psi_{L-2}\otimes s_{i,i+1}=((-1)^{n-1}\Psi_{L-2})\otimes (-s_{i,i+1})=(-1)^n \Psi_{L-2}\otimes s_{i,i+1}$. 
By uniqueness of the solution of \eqref{eigall},
we conclude that
\begin{equation}\label{prerecur}
\Psi_L(\ldots,\underset{i}{x\vphantom{\eta}},\underset{i+1}{x+2\eta},\ldots)
= \psi_i(x;\ldots) \Psi_{L-2}(\ldots)\otimes s_{i,i+1}
\end{equation}
where, by the same kind of argument as in previous sections,
$\psi_i$ is a theta function of its arguments of degree $1$ in the
$x_j$, $j\ne i,i+1$ and of degree $4n$ in $x$.


In order to fix the function $\psi_i$, we shall need the so-called wheel
condition vanishing relation. Let us first consider a special case of it:
suppose three successive spectral parameters $x_i,x_{i+1},x_{i+2}$
are of the form
$x,x+2\eta,x+4\eta$. Then according to Eq.~\eqref{prerecur} applied at $(i,i+1)$
and $(i+1,i+2)$,
\begin{multline*}
\mathcal{P}_{i,i+1}
\Psi_L(\ldots,x,x+2\eta,x+4\eta,\ldots)
=
\mathcal{P}_{i+1,i+2}
\Psi_L(\ldots,x,x+2\eta,x+4\eta,\ldots)
\\*
=
-
\Psi_L(\ldots,x,x+2\eta,x+4\eta,\ldots)
\end{multline*}
But the action of the symmetric group $\mathcal{S}_3$ on
$\mathbb{C}^2\otimes\mathbb{C}^2\otimes\mathbb{C}^2$ does not possess
the sign representation as a sub-representation; therefore
\[
\Psi_L(\ldots,x,x+2\eta,x+4\eta,\ldots)=0
\]
Now assume all other parameters $x_j$, $j\ne i,i+1,i+2$, are generic;
then according to Eq.~\eqref{unit}, $\check R(x_j-x_k)$ ($j\ne i,i+1,i+2$,
$k=i,i+1,i+2$) is an invertible operator.
Applying repeatedly the exchange relation \eqref{exch} to the equality above,
we conclude that
\begin{equation}\label{wheel}
\Psi_L(\ldots,x,\ldots,x+2\eta,\ldots,x+4\eta,\ldots)=0
\end{equation}
where the location of the three arguments is now arbitrary,
as long as the cyclic order is respected.
This is the general wheel condition
(the equality is true for
generic $x_j$, therefore for all $x_j$). 

Finally, using pseudo-periodicity relations \eqref{pseudoconj},
as well as flip relation \eqref{flip}, we conclude that
the wheel condition vanishing relation \eqref{wheel} is valid provided
the triplet of spectral parameters forms a wheel $x,x+2\eta,x+4\eta$
modulo $\pi,\pi\tau$ (not just $2\pi\tau$! a crucial technical point
which will be used repeatedly below).

We can now come back to our recurrence relation \eqref{prerecur}.
On the left hand side, we notice that as soon as one of the $x_j$,
$j\ne i,i+1$, is equal to $x-2\eta\pmod{\pi,\pi\tau}$, a wheel is formed
and $\Psi_L$ vanishes. Therefore
$\psi_i(x;\ldots)$ contains factors $\prod_{j(\ne i,i+1)}
\th_1(x-2\eta-x_j;p^2)\th_4(x-2\eta-x_j;p^2)$; 
moreover these exhaust its degree,
and noting that these factors can also be written up to a multiplicative
constant as $\th_1(x-2\eta-x_j;p)$,
we can rewrite Eq.~\eqref{prerecur}
\begin{equation}\label{recur}
\Psi_L(\ldots,\underset{i}{x\vphantom{\eta}},\underset{i+1}{x+2\eta},\ldots)
= cst
\prod_{j(\ne i,i+1)}\th_1(x-2\eta-x_j;p)
\,
 \Psi_{L-2}(\ldots)\otimes s_{i,i+1}
\end{equation}
More explicitly, it means that
\[
\Psi_{L;\alpha_1,\ldots,\alpha_L}
|_{x_{i+1}=x_i+2\eta}
=
\begin{cases}
0&\alpha_i=\alpha_{i+1}
\\
cst\, \alpha_i
\prod_{j(\ne i,i+1)}\th_1(x_i-2\eta-x_j;p)
\,
 \Psi_{L-2;\alpha_1,\ldots,\alpha_{i-1},\alpha_{i+2},\ldots,\alpha_L}
&\alpha_i\ne\alpha_{i+1}
\end{cases}
\]
The constant remains undetermined at this stage, since we have not
fixed the normalization of $\Psi_L$ yet.

A similar recurrence relation can be written for $x_{i+1}=x_i+2\eta+\pi\tau$
(the non-zero result occurring when $\alpha_i=\alpha_{i+1}$), but we shall
not need it.

\section{Partition function}
In the rest of this paper, we denote $\th(x):=\th_1(x;p)$
and $\th_k(x):=\th_k(x;p)$, $k=2,3,4$.
Since the contents of this section are not expected to generalize
outside $\eta=\pi/3$, we shall use $3\eta=0\pmod\pi$ to
replace $2\eta$ with $-\eta$ whenever possible.

\subsection{Definition}
We now introduce a quantity that naturally generalizes the squared norm
of the XYZ ground state (Eq.~\eqref{sqnorm}) to the inhomogeneous case:
\[
\Z_L(x_1,\ldots,x_L)=\braket{\Psi_L(-x_1,\ldots,-x_L)}{\Psi_L(x_1,\ldots,x_L)}
\]
where we have used the (real) scalar product:
$\braket{\Phi}{\Phi'}=\sum_{\alpha\in\{\uparrow,\downarrow\}^L}\Phi_\alpha \Phi'_\alpha$.

$\Z_L(x_1,\ldots,x_L)$ has the following interpretation: it is the ``partition
function'' of the inhomogeneous eight-vertex model on an infinite cylinder.
Indeed, assuming that we are in a regime of parameters where $\Psi_L$
is associated to the largest eigenvalue of the transfer matrix,
$\Psi_L(x_1,\ldots,x_L)$ corresponds to the partition function
on a half-infinite cylinder (pointing upwards) 
with given arrows at the boundary at the bottom.
A vertical mirror symmetry of the eight vertices correspond
in the weights \eqref{boltz} to $x\to -2\eta-x$ and a change of sign
of the weights $a$ and $b$, the latter being irrelevant with periodic boundary
conditions. So the partition function of the other half-infinite cylinder
(pointing downwards) is $\Psi_L(-x_1,\ldots,-x_L)$
($\Psi_L$ only depends on the differences of its arguments so the $-2\eta$
term is irrelevant).
We mean partition function in the following sense:
a one-point correlation function will be expressed as 
$\left< \mathcal{O}\right>=\frac{1}{Z}
\sum_{\alpha,\beta\in\{\uparrow,\downarrow\}^L}
 \Psi_{L;\alpha}(-x_1,\ldots,-x_L)\mathcal{O}_{\alpha,\beta}
\Psi_{L;\beta}(x_1,\ldots,x_L)$.

\subsection{Pseudo-periodicity}
According to its definition and \eqref{pseudoconjb}, 
$\Z_L$ is invariant by $x_i\to x_i+\pi$ for any given $i$.
Furthermore,
\begin{align*}
\Z_L(\ldots,x_i+\pi\tau,\ldots)&=
\sum_{\alpha\in\{\uparrow,\downarrow\}^L}
\braket{\Psi_{L}(\ldots,-x_i-\pi\tau,\ldots)}{\Psi_{L}(\ldots,x_i+\pi\tau,\ldots)}
\\
&=
(-p)^n e^{2n\mathrm{i}(-x_i-\pi\tau)+\mathrm{i}\sum_{j(\ne i)}x_j}
\bra{
\Psi_L(\ldots,-x_i,\ldots)}F_i
\eqbreak
(-p)^{-n}e^{-2n\mathrm{i}x_i+\mathrm{i}\sum_{j(\ne i)} x_j}
F_i\ket{\Psi_L(\ldots,x_i,\ldots)}
\qquad\text{by Eq.~\eqref{flip}}
\\
&=p^{-2n} z_i^{2n}\prod_{j(\ne i)} z_j^{-1}\, \Z_L(\ldots,x_i,\ldots)
\end{align*}
where it is reminded that $z_j=e^{-2\mathrm{i}x_j}$.

We reach the conclusion that $\Z_L$ is a theta function of degree $2n$ and
nome $p$ (as opposed to $p^2$ for $\Psi_L$) in each
variable $x_i$.

\subsection{Symmetry}
Given $i=1,\ldots,L-1$, we can use the exchange relation \eqref{exch}
and unitarity relation \eqref{unit} to write
\begin{alignat*}{2}
\Z_L(x_1,\ldots,&x_{i+1},x_i,\ldots,x_L)
\\
&=
\braket{%
\Psi_{L}(-x_1,\ldots,-x_{i+1},-x_i,\ldots,-x_L)}%
{\Psi_{L}(x_1,\ldots,x_{i+1},x_i,\ldots,x_L)}
\\
&=
\bra{
\frac{\check R_{i,i+1}(x_{i}-x_{i+1})}{r(x_i-x_{i+1})}
\Psi_L(-x_1,\ldots,-x_i,-x_{i+1},\ldots,-x_L)}
\eqbreak
\ket{\frac{\check R_{i,i+1}(x_{i+1}-x_{i})}{r(x_{i+1}-x_i)}
\Psi_L(x_1,\ldots,x_i,x_{i+1},\ldots,x_L)}
\\
&=
\bra{\Psi_L(-x_1,\ldots,-x_i,-x_{i+1},\ldots,-x_L)}
\frac{\check R_{i,i+1}(x_{i}-x_{i+1})}{r(x_i-x_{i+1})}
\eqbreak
\frac{\check R_{i,i+1}(x_{i+1}-x_{i})}{r(x_{i+1}-x_i)}
\ket{
\Psi_L(x_1,\ldots,x_i,x_{i+1},\ldots,x_L)}
\\
&=\Z_L(x_1,\ldots,x_L)
\end{alignat*}
where in the intermediate step we also used the fact that the $\check R$ matrix
is self-adjoint.
We conclude from this calculation that $\Z_L$ is a {\em symmetric}\/ function
of its arguments.

\subsection{Recurrence relation}
The recurrence relation \eqref{recur} for $\Psi_L$ implies
one for $\Z_L$:
\begin{align*}
\Z_L(\ldots,x,x+2\eta)
&=
\braket{\Psi_{L}(\ldots,x,x+2\eta)}{\Psi_{L}(\ldots,-x,-x-2\eta)}
\\
&\propto
\prod_{i=1}^{L-2} \th(x-2\eta-x_i) 
\braket{\Psi_{L-2}(\ldots)\otimes s_{L-1,L}}{\Psi_{L}(\ldots,-x,-x-2\eta)}
\\
&\propto
\prod_{i=1}^{L-2} \th(x-2\eta-x_i) 
\bra{\Psi_{L-2}(\ldots)\otimes s_{L-1,L}}P_{L-1,L}
\ket{\Psi_L(\ldots,-x,-x-2\eta)}
\\
&\propto
\prod_{i=1}^{L-2} \th(x-2\eta-x_i) 
\braket{\Psi_{L-2}(\ldots)\otimes s_{L-1,L}}{\Psi_{L}(\ldots,-x-2\eta,-x)}
\eqbreak
\text{by Eq.~\eqref{exch} with $x_{L-1}=-x$, $x_L=-x-2\eta$}
\\
&\propto
\prod_{i=1}^{L-2} \th(x-2\eta-x_i) \th((-x-2\eta)-2\eta-(-x_i))
\eqbreak
\braket{\Psi_{L-2}(x_1,\ldots,x_{L-2})\otimes s_{L-1,L}}%
{\Psi_{L-2}(-x_1,\ldots,-x_{L-2})\otimes s_{L-1,L}}
\\
&\propto
\prod_{i=1}^{L-2} \th^2(x-2\eta-x_i)
\,
\Z_{L-2}(\ldots)
\end{align*}
where $\propto$ means equal up to a multiplicative constant.
At this stage, we fix the normalization of $\Psi_L$ in such a way
that this constant disappears in the recurrence formula for $\Z_L$,
which becomes:
\begin{equation}\label{recZ}
\Z_L(\ldots,x,x+\eta)=\prod_{i=1}^{L-2}\th^2(x-\eta-x_i)\,\Z_{L-2}(\ldots)
\end{equation}
where we have also shifted $x\to x+\eta$ and used $3\eta=0\pmod\pi$.

Combined with the symmetry in its arguments,
the recurrence relation \eqref{recZ} satisfied by $\Z_L$
means that we can express
its specialization at $x_1=x_2\pm \eta,\ldots,x_L\pm \eta$
in terms of $\Z_{L-2}$. So we possess $4n$ values of $\Z_L$ as a function
of $x_1$; since it is a theta function of degree $2n$, these relations
are more than enough to determine $\Z_L$ inductively
(say, by Lagrange interpolation).

\subsection{Half-specialization}\label{halfspec}
At the moment, we do not know how to solve in a closed form the
recurrence relation above. However, note that we have twice as many
recurrence relations as needed to determine $\Z_L$. This suggests
to ``half-specialize'' $\Z_L$ in such a way that the number of recurrence
relations now matches the degree. 

Explicitly, assume $x_{i+n}=-x_i$, $i=1,\ldots,n$, and $x_L=0$.
After such a specialization, $\Z_L$ is an even function of $x_1,\ldots,x_n$,
and it has a {\em double zero}\/ at $x_i=\pm\eta$, $i=1,\ldots,n$.
Let us check the latter statement 
carefully. Since $\Z_L$ is a symmetric function of its
arguments, let us assume that we order them as
$x,0,-x\ldots$ (where $x$ is one of the $x_i$) 
and that we send $x$ to $-2\eta$ (which is equal to
$\eta$ modulo $\pi$). Then it is clear that $\Psi_L(x,0,-x,\ldots)$ forms
a wheel and therefore $Z_L(x,0,-x,\ldots)$ vanishes. 
However, $\Psi_L(-x,0,x,\ldots)$ does
not vanish, so that to show that the zero is double,
we need to go further. Write
\newcommand\der{\partial}
\begin{multline}\label{deriv}
\frac{\der}{\der x}
\Z_L(x,0,-x,\ldots)|_{x=-2\eta}
\\*
= \braket{\frac{\der}{\der x} \Psi_L(x,0,2\eta,\ldots)|_{x=-2\eta}
-\frac{\der}{\der x} \Psi_L(-2\eta,0,x,\ldots)|_{x=2\eta}}%
{\Psi_{L}(2\eta,0,-2\eta,\ldots)}
\end{multline}
Now apply recurrence relation \eqref{recur} to $\Psi_L(x,0,2\eta,\ldots)$;
we find that it is proportional to some vector at sites $j\ne 2,3$ tensor
$s_{2,3}$, and therefore the same its true of its derivative w.r.t.\ $x$
(one can be more explicit using $x=-2\eta$ but we shall not need it).
Similarly, $\Psi_L(-2\eta,0,x,\ldots)$ and its derivative w.r.t.\ $x$
are equal to $s_{1,2}$ tensor some vector at other sites.

On the other hand, applying the exchange relation \eqref{exch}
to $\Psi_L(2\eta,0,-2\eta)$
at $i=1$ implies that $P_{1,2}\Psi_L(2\eta,0,-2\eta,\ldots)\propto
\Psi_L(0,2\eta,-2\eta,\ldots)=0$ since a wheel is formed.
Similarly, the exchange relation at $i=2$ implies that $P_{2,3}\Psi_L(2\eta,0,-2\eta,\ldots)=0$.

We conclude that the expression \eqref{deriv} is zero by inserting
$P_{2,3}$ (resp.\ $P_{1,2}$) in the first (resp.\ second) term.
Therefore, taking into account evenness, we can write
\begin{equation}
\Z_L(x_1,\ldots,x_n,-x_1,\ldots,-x_n,0)
=
\prod_{i=1}^n \th^2(x_i-\eta)\th^2(x_i+\eta)
\,\X_n(x_1,\ldots,x_n)
\end{equation}
where $\X_n(x_1,\ldots,x_n)$ has the following properties,
as a direct consequence of the corresponding properties for $\Z_L$:
\begin{itemize}
\item $\X_n$ is a symmetric function of its arguments, and an even
theta function of degree $2(2n-1)$ in each.
\item It satisfies the recurrence relations:
\begin{align}
\X_n(\ldots,x,x+\eta)&=
\ph^2(x)\ph^2(x+\eta)\prod_{i=1}^{n-2} \th^4(x-\eta-x_i)\th^4(x-\eta+x_i)
\,\X_{n-2}(\ldots)
\\
\X_n(\ldots,\beta_k)&=\frac{\kappa_k}{\th_k^2(\eta)} 
\prod_{i=1}^{n-1} \th_k^4(x_i)\, \X_{n-1}(\ldots)
\qquad k=2,3,4
\end{align}
where $\ph(x)=\frac{\textstyle\th(2x)}{\textstyle\th(x)}=\frac{1}{\kappa}
\th_2(x)\th_3(x)\th_4(x)$,
$\kappa=\displaystyle\frac{\th_2\th_3\th_4}{2}=\th_2(\eta)\th_3(\eta)\th_4(\eta)$
and
$\kappa_2=1$, $\kappa_3=\kappa_4=-\nu^{2n-1}$ ($\nu=e^{-2\pi i/3}/\sqrt{p}$)
are pseudo-periodicity constants. 
\end{itemize}
where $\beta_{2,3,4}$ are representatives
of the three solutions of $2\beta_k+\eta=0\pmod {\pi,\pi\tau}$ excluding $\eta$,
namely, $\beta_2=\pi/2+\eta$, $\beta_3=\pi/2+\pi \tau/2+\eta$,
$\beta_4=\pi \tau/2+\eta$.

We now have at our disposal the specializations $x_1=\pm x_i\pm \eta,
\pm\beta_k$, $i=2,\ldots,n$, $k=2,3,4$, that is $4(n-1)+6=2(2n+1)$.
An even theta function of degree $2(2n-1)$ being determined by $2\times 2n$
values, we have enough recurrence relations to determine $\X_n$.

\subsection{Solution as Pfaffians}\label{sec:pfaff}
We first introduce the function:
\[
\A_2(x,y)=-\nu^2\frac{\th_2(\eta)}{\th_2(0)}\left(\th_3(x+\eta)\th_3(x-\eta)\th_4^2(y)
+
\th_4(x+\eta)\th_4(x-\eta)\th_3^2(y)\right)
\]
which has the following properties:
\begin{itemize}
\item It is symmetric function of $x,y$, and is an even theta
function of degree $2$ in each.
\item It satisfies the following recurrence relations:
\begin{align}\label{A2rec}
\A_2(x,x+\eta)&=-\nu^2\, \th_3(x)\th_3(x+\eta)\th_4(x)\th_4(x+\eta)
\\
\A_2(x,\beta_k)&=\nu^3\th_2^2(\eta)
\, \th_k^2(x)\qquad k=3,4
\label{A2rec2}
\end{align}
\end{itemize}

Next we claim the following: define
\begin{equation}
\A_n(x_1,\ldots,x_n)=
\prod_{1\le i<j\le n} \frac{\h(x_i,x_j)}{\th(x_i-x_j)\th(x_i+x_j)}
\ \Pf \M_n
\end{equation}
where 
\[
\h(x,y)=\th(\eta+x-y)\th(\eta+x+y)\th(\eta-x-y)\th(\eta-x+y)
\] 
and $\M_n$ is a skew-symmetric $2m\times 2m$ matrix, $m=\lceil n/2 \rceil$,
given by
\begin{equation}
(\M_n)_{ij}=
\begin{cases}
\f(x_i,x_j)&\text{$n$ even or $i,j<2m$}\\
-1&\text{$n$ odd, $i=2m$, $j<2m$}\\
1&\text{$n$ odd, $j=2m$, $i<2m$}\\
0&\text{$n$ odd, $i=j=2m$}
\end{cases}
\end{equation}
and
\begin{equation}
\f(x,y)
=\frac{\th(x-y)\th(x+y) \A_2(x,y)}
{\h(x,y)}
\end{equation}
Also define
\begin{equation}
\B_n(x_1,\ldots,x_n)=\A_{n+1}(x_1,\ldots,x_n,\beta_2)
\end{equation}
Then:
\begin{itemize}
\item $\A_{n}$ (resp.\ $\B_n$) 
is a symmetric function of its arguments, and an even
theta function of degree $2(n-1)$ (resp.\ $2n$) in each.
\item They satisfy the recurrence relations:
\begin{align}\label{recurA}
\A_n(\ldots,x,x+\eta)&=-\nu^2\, \th_3(x)\th_3(x+\eta)\th_4(x)\th_4(x+\eta)
\eqbreak
\prod_{i=1}^{n-2}\th^2(x-\eta-x_i)\th^2(x-\eta+x_i)\, \A_{n-2}(\ldots)
\\
\B_n(\ldots,x,x+\eta)&=-\nu^2\,
\th_2^2(x)\th_2^2(x+2\eta)
\th_3(x)\th_3(x+\eta)
\th_4(x)\th_4(x+\eta)
\eqbreak
\prod_{i=1}^{n-2}\th^2(x-\eta-x_i)\th^2(x-\eta+x_i)\, \B_{n-2}(\ldots)
\\
\A_n(\ldots,\beta_2)&=\B_{n-1}(\ldots)
\\
\B_n(\ldots,\beta_2)&=
-\nu^2\th_3^2(\eta)\th_4^2(\eta)
\prod_{i=1}^{n-1} \th_2^4(x_i) 
\, \A_{n-1}(\ldots)
\\
\A_n(\ldots,\beta_k)&=
\nu^n\th_2(\eta)(\nu\th_2(\eta))^{(-1)^n}
\prod_{i=1}^{n-1} \th_k^2(x_i) 
\,\A_{n-1}(\ldots)& k&=3,4
\\
\B_n(\ldots,\beta_k)&=
\nu^{n+1}\th_2(\eta)(\nu\th_2(\eta))^{-(-1)^n}
\frac{\th_3^2(\eta)\th_4^2(\eta)}{\th_k^2(\eta)}
\prod_{i=1}^{n-1} \th_k^2(x_i) 
 \,\B_{n-1}(\ldots)& k&=3,4
\end{align}
where conventionally $\A_0=\B_0=1$.
\end{itemize}

Let us show for example \eqref{recurA} for $n=2m$ even.
Assume that $x_{2m-1}=x$ approaches $x_{2m}+\eta=x'$. Then the matrix element $(\M_n)_{2m-1,2m}$
develops a pole: $(\M_n)_{2m-1,2m}\propto 1/(x-x')$;
and the other entries $(\M_n)_{ij}$ ($i<j$) remaining finite,
the only relevant contributions to the Pfaffian are those pairing $2m-1$ and $2m$, so we immediately have
\[
\A_n(\ldots,x,x+\eta) = \prod_{i=1}^{n-2} \frac{\h(x_i,x)\h(x_i,x+\eta)}{\th(x_i-x)\th(x_i-x-\eta)\th(x_i+x)\th(x_i+x+\eta)}
\A_2(x,x+\eta) \A_{n-2}(\ldots)
\]
where we have cancelled all factors in common to $\A_n$ and $\A_{n-2}$.

Now the remarkable phenomenon (using in a crucial way $\eta=\pi/3$) is that there are compensations in
the product, which simplifies to $\prod_{i=1}^{n-2} \th^2(x-\eta-x_i)\th^2(x-\eta+x_i)$. Finally,
we use Eq.~\eqref{A2rec} for $\A_2$ to reproduce the remaining prefactors on the r.h.s.\ of Eq.~\eqref{recurA}.

The other equations follow from similar reasonings.

Finally, we find that $\A_n \B_n$ satisfies all the recurrence relations of $\X_n$, or more precisely,
\[
\X_n(x_1,\ldots,x_n)=(-\nu^2 \kappa^2)^{-n}
\A_n(x_1,\ldots,x_n)\B_n(x_1,\ldots,x_n)
\]

\subsection{Further factorization as determinants}\label{sec:det}
Consider the following elliptic version of Tsuchiya's determinant
\cite{Tsuchiya,kup-symASM}:
(see a similar determinant in \cite{Fil-sos})
\begin{equation}\label{detH}
\H_{2m}(x_1,\ldots,x_m;x_{m+1},\ldots,x_{2m})=
\frac%
{\displaystyle
\prod_{i=1}^m \prod_{j=m+1}^{2m} 
\h(x_i,x_j)}
{\displaystyle\prod_{\substack{1\le i<j\le m\\\text{or}\\m+1\le i<j\le 2m}}
\th(x_i-x_j)\th(x_i+x_j)}
\det_{\substack{1\le i\le m\\ m+1\le j\le 2m}}\frac{1}{\h(x_i,x_j)}
\end{equation}
Conventionally, $\H_0=1$. Note that $\H_2=1$ as well.

\rem{technically
we don't need
the matrix to be elliptic, only that the pseudo-periodicity factors
in terms of the 2 variables}

The expression of $\H_{2m}$ has the disadvantage that it is only (apparently) symmetric in
the variables $\{x_1,\ldots,x_m\}$ and $\{x_{m+1},\ldots,x_{2m}\}$; in fact
we show in appendix~\ref{proofsym} that thanks to $\eta=\pi/3$,
it is indeed symmetric in all variables. 
In terms of each, it is an even theta function
of degree $2(m-1)$.

It is not hard to see that $\H_{2m}$ satisfies the following recurrence relation:
\begin{equation}\label{recH}
\H_{2m}(\ldots,x;\ldots,x+\eta)=\prod_{\substack{i=1,\ldots,m-1,\\m+1,\ldots,2m}}\th(x-\eta-x_i)\th(x-\eta+x_i) \H_{2m-2}(\ldots;\ldots)
\end{equation}
and the same if one exchanges $x$ and $x+\eta$.
These are the usual
recurrence relations satisfied by such determinants, as in the classical
case of the Izergin--Korepin determinant \cite{Kor,Iz-6v,ICK}, and similarly to the Pfaffians of Sect.~\ref{sec:pfaff}; 
the reasoning to derive Eq.~\eqref{recH} is identical
-- a pole develops in one of the entries of the determinant, 
reducing it to a determinant one size smaller. 

Now consider the function 
\[
\A'_{2m}(x_1,\ldots,x_{2m})=\H_{2m}(x_1,\ldots,x_m;x_{m+1},\ldots,x_{2m})\H_{2m+2}(x_1,\ldots,x_m,\beta_3;
x_{m+1},\ldots,x_{2m},\beta_4)
\]
$\A'_{2m}$ is an even theta function of its arguments,
of degree $2(2m-1)$, and using Eq.~\eqref{recH}, it satisfies the same recurrence relation
\eqref{recurA} as $\A_{2m}$,
for say $x_1=\pm x_j\pm\eta$, $j=m+1,\ldots,2m$. Thus the function is known at $2\times 2m$ values of $x_1$,
which determines it uniquely. Combined with $\A'_2=\A_2=1$,
we conclude by induction that $\A'_{2m}=\A_{2m}$. 


Similar arguments can be made for $\A_{2m-1}$ and $\B_n$. Together, we find
\begin{align}\label{AtoH}
\A_{2m}(x_1,\ldots,x_{2m})&=\H_{2m}(x_1,\ldots,x_{2m})\H_{2m+2}(x_1,\ldots,x_{2m},\beta_3,\beta_4)
\\
\A_{2m-1}(x_1,\ldots,x_{2m-1})&=\H_{2m}(x_1,\ldots,x_{2m-1},\beta_3)\H_{2m}(x_1,\ldots,x_{2m-1},\beta_4)
\\
\B_{2m}(x_1,\ldots,x_{2m})&=\H_{2m+2}(x_1,\ldots,x_{2m},\beta_2,\beta_3)
\H_{2m+2}(x_1,\ldots,x_{2m},\beta_2,\beta_4)
\\
\B_{2m-1}(x_1,\ldots,x_{2m-1})&=\H_{2m}(x_1,\ldots,x_{2m-1},\beta_2)
\H_{2m+2}(x_1,\ldots,x_{2m-1},\beta_2,\beta_3,\beta_4)
\label{BtoH}
\end{align}
These are the only 8 possible
specializations at $\beta_{2,3,4}$,
corresponding to subsets of
$\{\beta_2,\beta_3,\beta_4\}$,
since applying any such specialization twice amounts to
the shift $n\to n-2$.

\subsection{Alternative determinant formula}
Here we follow the same general method as in \cite{Strog-IK} (see also appendix B of \cite{artic31}). 
Define 
\[
\g(x,y)=\frac{\th(2x)\th(2y)}{\h(x,y)}
\]
$\g(x,y)$ is an odd elliptic function of $x$ and $y$. 
One further observes that
for all $x,y$,
\[
\g(x,y)+\g(x+\eta,y)+\g(x+2\eta,y)=0
\]
and similarly for $y$.
Therefore, $\det_{i,j}\g(x_i,x_j)$, as a function of any of its arguments,
satisfies the same three-term relation. Now define
\begin{align*}
\S_{2m}(x_1,\ldots,x_{2m})&=\prod_{1\le i<j\le 2m} \th(x_i-x_j)
\prod_{1\le i\le j\le 2m} \th(x_i+x_j) 
\ \H_{2m}(x_1,\ldots,x_{2m})
\\
&=\prod_{\substack{1\le i<j\le m\\\text{or}\\m+1\le i<j\le 2m}}\prod_{k=0}^2 \th(x_i-x_j+k\eta)\th(x_i+x_j+k\eta)
\det_{\substack{1\le i\le m\\ m+1\le j\le 2m}}\g(x_i,x_j)
\end{align*}
Since the prefactor is invariant by $x_i\to x_i+\eta$ for any $i$, we have the same three-term relation
for $\S_{2m}$. In summary:
\begin{itemize}
\item $\S_{2m}$ is a skew-symmetric function of its arguments $x_i$,
and an odd theta function of degree $6m$ in each.
\item It satisfies
\[
\S_{2m}(\ldots,x,\ldots)+
\S_{2m}(\ldots,x+\eta,\ldots)+
\S_{2m}(\ldots,x+2\eta,\ldots)=0
\]
\end{itemize}

The space of odd theta functions of degree $6m$ is of dimension
$3m$, a possible basis being
\[
s_k(x)=e^{2ikx}\th_3(k\pi\tau+6mx,p^{6m})-e^{-2ikx}\th_3(k\pi\tau-6mx,p^{6m})
\qquad k=0,\ldots,3m-1
\]
$s_k(x)$ satisfies the relation $s_k(x)+s_k(x+\eta)+s_k(x+2\eta)=0$ iff $k\ne 0\pmod 3$.
The sequence $(1,2,4,5,\ldots,3m-2,3m-1)=(k_1,\ldots,k_{2m})$ is of cardinality $2m$,
which is the number of variables of $\S_{2m}$, so we conclude that $\S_{2m}$ is proportional to the ``Slater
determinant''
\begin{equation}\label{slater}
\S_{2m}(x_1,\ldots,x_{2m})\propto \det_{i,j=1,\ldots, 2m} s_{k_j}(x_i)
\end{equation}
We shall not need the proportionality constant, only that it is nonzero
(for generic $p$). 

\rem{Fay identity? i.e., fact that denominator of $H$ is essentially
same determinant $s_k(x_i)$}

\subsection{Uniformization}
Although the formulae above are simple to derive, they are a bit too cumbersome
to be used, especially in the homogeneous limit. Since all functions
we consider are theta functions of definite parity, there is a rational uniformization,
and we use from now on the following parameterization:
\[
w(x)=(1-\zeta^2)^{-1/3}\frac{\th^2(x)}{\th(x-\eta)\th(x+\eta)}
\]
In terms of the original Boltzmann weights \eqref{boltz},
we have $(1-\zeta^2)w(x)=\frac{(a(x-\eta)+b(x-\eta))^2}{a(x-\eta) b(x-\eta)}$. 

Note the special values
\begin{align*}
w(\beta_2)&=-\frac{1}{2}=J_2 & w(\beta_3)&=\frac{1}{1+\zeta}=J_3 & w(\beta_4)&=\frac{1}{1-\zeta}=J_4
\end{align*}
which explains the labelling we have chosen for the coupling constants
$J_{2,3,4}$.

This parameterization
has the advantage that the wheel condition becomes simple to express:
three spectral parameters form a ``wheel'' $\pm x,\pm(x+\eta),\pm(x+2\eta)$
iff the corresponding variables $w,w',w''$ satisfy
\begin{equation}
\left\{
\begin{aligned}
w+w'+w''&=\frac{3+\zeta^2}{1-\zeta^2}
\\
w\, w' w''&=\frac{1}{1-\zeta^2}
\end{aligned}
\right.
\end{equation}
and therefore, two parameters form a ``2-string'' $\pm x,\pm(x+\eta)$ iff
the corresponding variables $w,w'$ satisfy $h(w,w')=0$, where
\begin{equation}
h(w,w')=1-(3+\zeta^2)ww'+(1-\zeta^2)ww'(w+w')
\end{equation}
This formula allows to rewrite the recurrence formulae in this new
parameterization, but due to the fact that it is quadratic in $w$ and $w'$,
the result is somewhat cumbersome and we shall not write it explicitly. 

We also redefine the functions by dividing them by a ``reference'' even theta function of degree 2 to the appropriate power, 
here $\th(x-\eta)\th(x+\eta)$,
and absorbing some constants in the normalization.
That is, we define 
\begin{align*}
A_n(w(x_1),\ldots,w(x_n))&=a_n\,\frac{\A_n(x_1,\ldots,x_n)}{\prod_{i=1}^n
(\th(x_i-\eta)\th(x_i+\eta))^{n-1}}
\\
B_n(w(x_1),\ldots,w(x_n))&=b_n\frac{\B_n(x_1,\ldots,x_n)}{\prod_{i=1}^n
(\th(x_i-\eta)\th(x_i+\eta))^{n-1}}
\end{align*}
where $a_n$ and $b_n$ are constants which are implicitly defined by the expressions below, and whose
explicit expression we shall not need.

In particular,
\[
A_2(w,w')=
ww'-(w+w')+\frac{1+\zeta^2}{1-\zeta^2}
\]
and if we define
\[
 f(w,w')=\frac{(w-w')A_2(w,w')}{h(w,w')}
\]
which is such that $f(w(x),w(y))=-\zeta q/\sqrt{p}\ \f(x,y)$,
then we have:
\begin{equation}\label{defA}
A_{n}(w_1,\ldots,w_n)=\prod_{1\le i<j\le n}
\frac{h(w_i,w_j)}{w_i-w_j}
\ \Pf M_n
\end{equation}
where $M_n$ is identical to $\M_n$, except entries $\f(x_i,x_j)$ are replaced
with entries $f(w_i,w_j)$;
and
\[
B_n(w_1,\ldots,w_n)=A_{n+1}(w_1,\ldots,w_n,J_2)
\]
as well as
\[
X_n(w_1,\ldots,w_n)=2^{n+1} A_n(w_1,\ldots,w_n)B_n(w_1,\ldots,w_n)
\]
where the numerical coefficient has been adjusted so that in the rational limit,
the normalization of $X_n$ coincides with the one discussed in Sect.~\ref{sec:homlim}.

A further advantage of this new normalization is that $A_n$ and $B_n$ are polynomials
in $w_1,\ldots,w_n$ and also of $\zeta$, up to a conventional denominator in powers of $1-\zeta^2$
which we have added for convenience.

Similarly, we can define
\[
H_{2m}(w_1,\ldots,w_{2m})=
\frac%
{\displaystyle
\prod_{i=1}^m \prod_{j=m+1}^{2m} 
h(w_i,w_j)}
{\displaystyle\prod_{\substack{1\le i<j\le m\\\text{or}\\m+1\le i<j\le 2m}}
(w_i-w_j)}
\det_{\substack{1\le i\le m\\ m+1\le j\le 2m}}\frac{1}{h(w_i,w_j)}
\]
and then the relations (\ref{AtoH}--\ref{BtoH}) expressing $\A,\B$ in terms of $\H$ remain the same;
more compactly, one can write:
\[
X_n(\ldots)=2^{n+1} \prod_{\substack{S\subset \{J_2,J_3,J_4\}\\
|S|=n\pmod 2}} H_{n+|S|}(\ldots,S)
\]



There are various alternative formulae, for example
\[
H_{2m+2}(w_1,\ldots,w_{2m},J_3,J_4)=
\frac%
{\displaystyle\prod_{i=1}^m \prod_{j=m+1}^{2m} 
h(w_i,w_j)}
{\displaystyle
\prod_{\substack{1\le i<j\le m\\\text{or}\\m+1\le i<j\le 2m}}(w_i-w_j)}
\det_{\substack{1\le i\le m\\ m+1\le j\le 2m}}\frac{A_2(w_i,w_j)}{h(w_i,w_j)}
\]

Finally, the transformations
\[
\zeta\to -\zeta \qquad \zeta\to \frac{\zeta+3}{\zeta-1}
\]
generate the group of permutations of the three coupling constants
$J_2,J_3,J_4$. Via the uniformization $w(x)=w_\zeta(x)$, this translates into
the symmetry of permutations of non-trivial solutions of $2x+\eta=0$.
The function $h(w,w')=h_\zeta(w,w')$ 
itself possesses this symmetry, in the sense that
\[
h_\zeta(w,w')=h_{-\zeta}(w,w')
\qquad
h_\zeta(w,w')=h_{(\zeta+3)/(\zeta-1)}
\Big(\frac{\zeta-1}{2}w,\frac{\zeta-1}{2}w'\Big)
\]
which is consistent with $w_\zeta(\beta_{3,4})=w_{-\zeta}(\beta_{4,3})$ and
\begin{align*}
w_{(\zeta+3)/(\zeta-1)}(\beta_2)&=-1/2=\frac{\zeta-1}{2} J_4\\
w_{(\zeta+3)/(\zeta-1)}(\beta_3)&=\frac{\zeta-1}{2(\zeta+1)}=\frac{\zeta-1}{2} J_3\\
w_{(\zeta+3)/(\zeta-1)}(\beta_4)&=\frac{1-\zeta}{4}=\frac{\zeta-1}{2} J_2\\
\end{align*}

\section{Homogeneous limit of the partition function}
\subsection{Summary}\label{summary}
The homogeneous limit is obtained by setting all spectral parameters equal;
in the half-specialized partition function $\X_n$, this is achieved
by sending all $x_i$ to zero.


In this section, we use the following notation: we omit parameters
that are set to zero, e.g., $H_{2m}=H_{2m}(\underbrace{0,\ldots,0}_{2m})$.
This is unambiguous because the total number of variables is given in subscript.
Here are some values of $H_{2m}$ for $m=0,1,2,3$:
\begin{align*}
H_{2m}&=1,&&1,&&3+\zeta^2,&&26 + 29 \zeta^2 + 8 \zeta^4 + \zeta^6
\\
2^{m-1} H_{2m}(J_2)&=&&1,&&7+\zeta^2,&&143 + 99 \zeta^2 + 13 \zeta^4 + \zeta^6
\\
H_{2m}(J_3)&=&&1,&&2+\zeta+\zeta^2,&&11 + 12 \zeta + 21 \zeta^2 + 10 \zeta^3 + 7 \zeta^4 + 2 \zeta^5 + \zeta^6
\\
H_{2m}(J_4)&=&&1,&&2-\zeta+\zeta^2,&&11 - 12 \zeta + 21 \zeta^2 - 10 \zeta^3 + 7 \zeta^4 - 2 \zeta^5 + \zeta^6 
\\
2^{m-1} H_{2m}(J_2,J_3)&=&&1,&&5+2\zeta+\zeta^2,&&
66 + 63 \zeta + 81 \zeta^2 +  30 \zeta^3 + 12 \zeta^4 + 3 \zeta^5 + \zeta^6
\\
2^{m-1} H_{2m}(J_2,J_4)&=&&1,&&5-2\zeta+\zeta^2,&&
66 - 63 \zeta + 81 \zeta^2 -  30 \zeta^3 + 12 \zeta^4 - 3 \zeta^5 + \zeta^6
\\
H_{2m}(J_3,J_4)&=&&1,&&1+\zeta^2,&&3+9\zeta^2+3\zeta^4+\zeta^6
\\
2^{m-1}H_{2m}(J_2,J_3,J_4)&=&&&&3+\zeta^2,&&21+39\zeta^2+3\zeta^4+\zeta^6
\end{align*}
We recognize the reciprocal polynomials of those
occurring in conjecture E of \cite{BM-P6b}: the correspondence of notations
is that for $m\ge 1$,
$H_{2m}=\zeta^{2m(m-1)} q_{m-1}(1/\zeta)$,
$H_{2m}(J_3)=\zeta^{2m(m-1)} p_{m-1}(1/\zeta)$, 
$2^{m-1}H_{2m}(J_2,J_4)=\zeta^{2m(m-1)}p_{-m}(1/\zeta)$, 
$2^{m-1}H_{2m}(J_2,J_3,J_4)=\zeta^{2m(m-1)} q_{-m}(1/\zeta)$.
All other sequences can be obtained by permutations of the $\{ J_2, J_3, J_4\}$,
and can therefore be obtained by iterating the transformations
$\zeta\to -\zeta$ and $\zeta\to \frac{\zeta+3}{\zeta-1}$, as explained
at the end of last section.
All the properties listed in conjecture E of \cite{BM-P6b}
can thus be checked on the $H_{2m}$.

If we recombine the $H_{2m}$ in pairs to form $A_n$ and $B_n$,
we recognize the reciprocal polynomials of the $s_n$ of
\cite{BM-P6b}: (see their appendix A)
\begin{align}
A_n&=\zeta^{2\lfloor n^2/4\rfloor}s_n(1/\zeta^2)& n&\ge 0
\\
B_n&=(2/3)^n \zeta^{2\lfloor(n+1)^2/4\rfloor}s_{-n-1}(1/\zeta^2)&n&\ge 0
\end{align}
from which we conclude
\[
X_n = 2^{n+1}  A_n  B_n
=2(4/3)^n \zeta^{n(n+1)} s_n(1/\zeta^2) s_{-n-1}(1/\zeta^2)
\]
which coincides with the expression given in Conjecture 1 of \cite{BM-P6b}
up to the factor of two (which is due to our slightly different
way of lifting the two-fold degeneracy: $F_\ast \Psi_L=(-1)^n\Psi_L$
effectively duplicates every entry of $\Psi_L$ compared to \cite{BM-P6b}).

See also appendix~\ref{zetazero} 
for an explanation of the constant terms of the various
polynomials above.

In the rest of this section, it is convenient to denote $\alpha=1-\zeta^2$.
We shall show that the various
polynomials above satisfy (differential) recurrence relations.

\subsection{Linear relations}
We first derive certain linear relations satisfied by $H_{2m}(w_1,\ldots,w_{2m})$. We shall need them
to relate the various derivatives of $H_{2m}$ at $w_i=0$.

Define
\begin{equation}\label{defD}
D_{2m}(w_1,\ldots,w_m;w_{m+1},\ldots,w_{2m})
=
\det_{\substack{1\le i\le m\\ m+1\le j\le 2m}}g(w_i,w_j)
\end{equation}
with $g(u,v)=\frac{1}{h(u,v)}$; we recall that $h(u,v)=1+uv(\alpha(u+v+1)-4)$.
In other words,
\begin{equation}
\label{HtoD}
H_{2m}(w_1,\ldots,w_{2m})=
\frac%
{\displaystyle
\prod_{i=1}^m \prod_{j=m+1}^{2m} 
h(w_i,w_j)}
{\displaystyle\prod_{\substack{1\le i<j\le m\\\text{or}\\m+1\le i<j\le 2m}}
(w_i-w_j)}
D_{2m}(w_1,\ldots,w_m;w_{m+1},\ldots,w_{2m})
\end{equation}

Also define
\begin{equation}\label{StoD}
S_{2m}(w_1,\ldots,w_{2m})=\prod_{1\le i<j\le 2m}(w_i-w_j) H_{2m}(w_1,\ldots,w_{2m})
\end{equation}

\subsubsection{A first order differential/divided difference equation}
We start from the following identity, which can be checked directly:
\begin{multline*}
\rho(u)
\der_u
g(u,v)
+
\rho(v)
\der_v g(u,v)
\\*
+
2(1-\alpha)(8+\alpha)\der_\alpha g(u,v)
+
\left(
\sigma(u)+\sigma(v)
\right)g(u,v)+(\delta_u+\delta_v) g(u,v)=0
\end{multline*}
where $\rho(u)=(1+2u)(4-6u+u\alpha+u^2\alpha)$, $\sigma(u)=5u(\alpha-4+4\alpha u)$,
$\der_u$ is the usual partial derivative $\frac{\der}{\der u}$, 
and $\delta_u$ is the {\em divided difference}\/ operator: $\delta_u \phi(u)=\frac{\phi(u)-\phi(0)}{u}$
for any function $\phi(u)$.

Then, one can easily prove starting from \eqref{defD} 
(for example by writing $D_{2m}$ as a sum over permutations and grouping
together the summands for values of the index connected by the permutation)
\begin{multline}\label{diveq}
\left(\sum_{i=1}^{2m} \left(\rho(w_i)\der_{w_i}+\sigma(w_i)+\delta_{w_i}\right)
+2(1-\alpha)(8+\alpha)\der_\alpha\right)
\\
D_{2m}(w_1,\ldots,w_m;w_{m+1},\ldots,w_{2m})
=0
\end{multline}

In principle, by using relation \eqref{HtoD}, one can reformulate this identity in terms of
$H_{2m}$, but the result is not particularly illuminating and we shall not need it.

\rem{one could rewrite slightly more easily the $\alpha$ derivative by introducing
$9/\zeta^2-1$ as variable... even better variable is $\alpha=\xi-4+4/\xi$}

\subsubsection{A second order differential equation}
Starting from the differential equation satisfied by $\th_3$, namely,
$(\frac{\der^2}{\der x^2}+4\,p\frac{\der}{\der p})\th_3(x;p)=0$, we find
\[
\left(\frac{\der^2}{\der x^2}+24m\, p\frac{\der}{\der p}\right)s_k=k^2 s_k
\]
According to Eq.~\eqref{slater}, this implies that
\begin{equation}\label{prediffeqa}
\left(\sum_{i=1}^{2m} \frac{\der^2}{\der x_i^2}+24m \,p\frac{\der}{\der p}\right) \S_{2m}(x_1,\ldots,x_{2m})=
c_m
\S_{2m}(x_1,\ldots,x_{2m})
\end{equation}
where $c_m$ is $m(6m^2-1)$ plus some $p$-dependent constant related to
the normalization of $\S_{2m}$.

After switching to our rational parameterization and from $p$ to $\alpha$, 
we find the following equation for $S_{2m}$:
\begin{equation}\label{prediffeq}
\left(
\sum_{i=1}^{2m} (\gamma_2(w_i) \der_i^2+\gamma_1(w_i)\der_i+\gamma_0(w_i))+24m\alpha(1-\alpha)(8+\alpha)\der_\alpha
\right) S_{2m}(w_1,\ldots,w_{2m})=0
\end{equation}
where the coefficients are entirely determined
except the constant term of $\gamma_0(w)$. The latter is determined
by the large $w=(w_1,\ldots,w_{2m})$ expansion: from (\ref{defD}--\ref{StoD}) 
one easily derives
\[
S_{2m}(w_1,\ldots,w_{2m})=\alpha^{m(m-1)}\prod_{i=1}^{2m}w_i^{m-1}
\prod_{1\le i<j\le 2m}(w_i-w_j)\ (1+O(w^{-3}))
\]
and expanding \eqref{prediffeq} up to second subleading order 
fixes the constant.
We find the rather unpleasant expressions:
\begin{align*}
\gamma_0(w)&=
18\alpha^2(m-1)(3m-2)w^2
+6\alpha(\alpha-4)(3m-2)(4m-3)w\qquad
\eqbreak
+64 m^2-12 \alpha ^2 m+96 \alpha 
   m-192 m+80
+5 \alpha ^2-40 \alpha +10 \alpha ^2 m^2-20 \alpha  m^2
\\
\gamma_1(w)&=-36\alpha^2(m-1)w^3
-6\alpha(\alpha-4)(10m-9)w^2
\eqbreak +6(3 \alpha ^2-24 \alpha -4 \alpha ^2 m+12 \alpha  m-32 m+48)w-36\alpha
\\
\gamma_2(w)&=6w (\alpha  w-4) (\alpha +\alpha  w^2+2 \alpha  w-4 w)
\end{align*}

\subsubsection{Homogeneous limit}
We now take the homogeneous limit in two steps: we first send $w_1,\ldots,w_m$ to $u$
and $w_{m+1},\ldots,w_{2m}$ to $v$ and then expand around $u,v=0$.
$H_{2m}$, being a symmetric function of $w_1,\ldots,w_{2m}$, only has one independent first
derivative (resp.\ two independent second derivatives), which with our specialization correspond to
$\frac{\der}{\der u} H_{2m}=\frac{\der}{\der v} H_{2m}$ (resp.\ 
$\frac{\der^2}{\der u^2} H_{2m}=\frac{\der^2}{\der v^2} H_{2m}$
and $\frac{\der^2}{\der u\der v}H_{2m}$). 

Taking this limit in Eqs.~\eqref{diveq} and \eqref{prediffeq} is a rather tedious procedure
which we shall not describe in detail. Expanding to first non-trivial order these equations produces
the same result, namely the first equation below. This equation is a first order differential
equation, and so we can differentiate it once w.r.t.\ $\alpha$, resulting in a second order equation
(second equation below).
Expanding to the next order Eqs.~\eqref{diveq} and \eqref{prediffeq} produces two {\em distinct}\/ second order differential equations. Finally, we find:
\begin{multline}\label{bigmess}
\left(
\vcenter{\halign{&\hfil$\scriptscriptstyle #$\hfil\crcr
0 & 0 & 0 & 0 & 2(4 m+1) & 2m (1-\alpha) (\alpha +8) & m^2(m-1)(2+\alpha) \cr
 0 & 0 & 2m (1-\alpha ) (\alpha +8) & 2(4 m+1) & 0 & m (\alpha  m^2+2 m^2-\alpha  m-2 m-4 \alpha -14) & m^2(m-1) \cr
 2(2 m+1) & 4 m & 0 & 2m (1-\alpha) (\alpha +8) &\, m (m^2-m+1) (2+\alpha) & 0 & m^2(m-1) (4-\alpha) \cr
 2 (4 m^2+m-2) \alpha  &\, -2 m (4 m+1) \alpha  & 0 & 0 & m (2 m+1) (\alpha -4)^2 & 0 & -(m-1) m^2 (2 m+1) (\alpha -4) \alpha  \cr
}}
\right)
\\*
\left(
\begin{smallmatrix}
\frac{\der^2}{\der u\der v} H_{2m}
\\
\frac{\der^2}{\der u^2} H_{2m}
\\
\frac{\der^2}{\der \alpha^2} H_{2m}
\\
\frac{\der^2}{\der u\der \alpha} H_{2m}
\\
\frac{\der}{\der u} H_{2m}
\\
\frac{\der}{\der \alpha} H_{2m}
\\
H_{2m}
\end{smallmatrix}
\right)
=0
\end{multline}

There are 4 relations for seven derivatives, so they can all be expressed
in terms of derivatives w.r.t.\ $\alpha$ only.

A similar reasoning can be made when all variables are specialized to $0$
except one, or two, or three, are specialized to a subset of $\{J_2, J_3,J_4\}$.
The result is given in appendix~\ref{moremesses}.

\subsection{Bilinar recurrence relations}
We now show how to derive differential bilinear recurrence relations
for $H_{2m}$ and its variants. In fact these relations were mentioned, but
not written explicitly, in paragraph 3 of \cite{BM-P6b}.

Similarly to the previous paragraph, we first consider the quantity
\[
H_{2m}(\underbrace{u,\ldots,u}_m,\underbrace{v,\ldots,v}_m)
=g(u,v)^{-m^2}
\det_{0\le i,j\le m-1} 
\left(
\frac{1}{i!j!}\frac{\der^{i+j}}{\der u^i \der v^j} g(u,v)
\right)
\]

A standard application of the Jacobi--Desnanot identity (see \cite{artic12} for
the simpler case of the Izergin--Korepin determinant) to the determinant
in the right hand side produces
the Toda lattice equation:
\[
\frac{1}{m^2}\frac{\der^2}{\der u\der v}
\log H_{2m}(u,\ldots,v,\ldots)=
- \frac{\der^2}{\der u\der v}\log g(u,v)+
\frac{H_{2(m+1)}(u,\ldots,v,\ldots)H_{2(m-1)}(u,\ldots,v,\ldots)}{g(u,v)^2 H_{2m}(u,\ldots,v,\ldots)^2}
\]
The left hand side involves first and second derivative of $H_{2m}(u,\ldots,v,\ldots)$, which at $u=v=0$ can be reexpressed in terms of derivatives w.r.t.\ $\alpha$ thanks
to Eq.~\eqref{bigmess}. The result is:
\begin{equation}\label{diffeq}
C_0\, H_{2(m+1)}H_{2(m-1)}=
C_1\,H_{2m} H''_{2m} -C_2\, (H'_{2m})^2
+C_3\,H_{2m} H'_{2m}+C_4\, H_{2m}^2 
\end{equation}
where all derivatives are w.r.t.\ $\alpha$, and
\begin{align*}
C_0
&=4 \alpha  (4 m-1) (4 m+1)^2 (4
   m+3)
\\
C_1
&=4 (\alpha -1)^2 \alpha  (\alpha +8)^2 (4 m+1)^2\\
C_2
&=4 (\alpha -1)^2 \alpha  (\alpha +8)^2 (4 m-1) (4 m+3)\\
C_3
&=2 (\alpha -1) (\alpha +8)
   (\alpha ^2+28 \alpha +24 \alpha ^2 m^2+304 \alpha  m^2-256 m^2
+20 \alpha ^2 m+208 \alpha  m-192 m-32)\\
C_4
&=6 \alpha ^2-24
   \alpha +4 \alpha ^3 m^4-1008 \alpha ^2 m^4+3408 \alpha  m^4+512 m^4-4 \alpha ^3 m^3-984 \alpha ^2 m^3
+4032 \alpha  m^3
\eqbreak
-128 m^3
+\alpha
   ^3 m^2-142 \alpha ^2 m^2+1028 \alpha  m^2-320 m^2-\alpha ^3 m+28 \alpha ^2 m-44 \alpha  m-64 m\\
 \end{align*}
Note that contrary to Eq.~\eqref{bigmess}, Eq.~\eqref{diffeq} is a closed relation
allowing to compute inductively the $H_{2m}$ as polynomials of $\alpha=1-\zeta^2$.

A similar computation produces differential recurrence relations of the same
form for the other factors of $X_n$. The coefficients are given in 
appendix~\ref{moremesses}.
Together, they allow to compute the full squared norm
$X_n$ inductively.

\section{Conclusion and prospects}
In this paper, we have considered the inhomogeneous eight-vertex model
with periodic boundary
conditions in odd size and crossing parameter $\eta=\pi/3$. We have provided
a basic setup for the computation
of the fully inhomogeneous generalization of the ground state eigenvector
of the XYZ spin chain, and then went on to compute the partition function
on an infinite cylinder, which generalizes the squared norm of the ground
state eigenvector, when the spectral parameters are ``half-specialized'',
i.e., form pairs $x,-x$. We have provided a variety of explicit
expressions for this partition function in terms of Pfaffians and
determinants. Interestingly, one can then obtain self-contained expressions
in the homogeneous limit for the squared norm, without any more reference
to the inhomogeneous case, by allowing differentiation w.r.t.\ the
variable parameterizing the line $\eta=\pi/3$ (elliptic nome, or $\zeta$).
These expressions take the form of bilinear differential recurrence relations
(cf Eq.~\eqref{diffeq}).

In order to derive such differential relations, we have used certain
differential (and divided difference) relations satisfied by the
inhomogeneous partition function. In fact,
we have strictly limited ourselves to the relations that were needed
for our purposes, but it seems that this 
is only the tip of the iceberg: one should investigate in more detail
the structure of the set of such equations. It would be interesting
to understand the role of the full symmetry of arguments of
the Izergin--Korepin type determinant \eqref{detH}.

Note that we have not been able to obtain an expression for the fully
inhomogeneous partition function, but if we compare to the work of
Rosengren for the 8VSOS model  \cite{rosengrenb} there is also
no simple expression for the fully inhomogeneous partition function.
Inversely, it would be interesting to see if the ``half-specialization''
trick helps in this context. More generally, as noted in \cite{BM-P6b},
there are many ressemblances between the work \cite{rosengrenb} and our
present setup, which should be clarified.

Another connection which should be more thoroughly 
explored is with the supersymmetric
models of lattice fermions of \cite{FH-susy,FH-susy2}.

It is clear that the present methods should allow to compute
more quantities such as individual entries of the ground state,
or certain correlation functions (see the recent work
\cite{Cantini-EFP} in the XXZ setting).

Some more directions which should be explored are: the relation
to the quantum Knizhnik--Zamolodchikov--Bernard ($q$KZB) equation
and to the $q$KZB heat equation \cite{FV-qKZB}, which should be the
right framework for part of section \ref{pties}, especially in view of
a generalization to arbitrary $\eta$;
the connection to nonsymmetric elliptic Macdonald polynomials;
\rem{cf math/0309452}
the use of matrix model techniques
to analyze the determinants of Izergin--Korepin type found here,
as in \cite{artic13}; and the meaning of the 
connection to the Painlev\'e VI equation,
which is emphasized in \cite{BM-P6a,BM-P6b}.

Finally, it would be interesting to find a combinatorial interpretation
for the (positive integer) entries 
of the polynomials of section \ref{summary}, beyond their value at $\zeta=0$.

\rem{relation to qkzb equation? qkzb ``heat'' (as in theta fns satisfying heat)? differential vs difference? why doesnt qkzb have any nome variation?
direct diff eq satisfied by $\Psi$?}

\rem{relation to enumeration problems, DPPs etc... quadratic in denominator
issue of quadratic denominator... if denominator was linear would be trivial (look at my polynomials)}

\appendix
\section{The $\zeta\to 0$ trigonometric limit}\label{zetazero}
The trigonometric limit is obtained by sending $\zeta$ to $0$.
The Boltzmann weights \eqref{boltz} of the eight-vertex model turn into those
of the six-vertex (the weight $d$ go to zero). In this limit
the results of this paper should be closely related to the computations
of \cite{artic36}. Note that the ``quadratic'' sum rule considered here
was actually not computed in \cite{artic36} -- instead
the quantity $\sum_\alpha \Psi_\alpha(z_1,\ldots,z_L)^2$ was used there.
However, the same argument of degeneracy of the scalar product allows
to conclude that
\begin{align}\label{Ztrig}
\Z_L(x_1,\ldots,x_L)&=3^{-n}
(\sum_\alpha \Psi_\alpha(x_1,\ldots,x_L))
(\sum_\alpha \Psi_\alpha(-x_1,\ldots,-x_L))
\\
&=3^{-n^2}
s_{Y_L}(z_1,\ldots,z_L) s_{Y_L}(z_1^{-1},\ldots,z_L^{-1})\notag
\end{align}
where $s_\lambda$ is the Schur function associated to partition $\lambda$,
and $Y_L=(\lfloor (L-i)/2\rfloor)_{i=1,\ldots,L}$.
In the homogeneous limit, $s_{Y_n}(1,\ldots,1)=3^{n(n-1)/2}
\prod_{j=1}^n \frac{(3j)!(j-1)!}{(2j)!(2j-1)!}$
and together we have $\Z_L=A_{HT}(L)$, where $A_{HT}(L)=1,3,25,588\ldots$ 
is the number
of Half-Turn Symmetric Alternating Sign Matrices \cite{kup-symASM,RS-halfturn}.

The half-specialization of section~\ref{halfspec} produces the following
factorization:
\begin{equation}\label{halfspectrig}
s_{Y_L}(1,z_1,z_1^{-1},\ldots,z_n,z_n^{-1})=\prod_{i=1}^n (1+z_i+z_i^{-1})\,
\chi_{Y_n}(z_1,\ldots,z_n)
\chi_{Y_{n+1}}(z_1,\ldots,z_n,\omega)
\end{equation}
where $\chi_\lambda$ is the symplectic character, defined by:
$\chi_\lambda(z_1,\ldots,z_n)=\frac{\det(z_i^{\lambda_j+n-j+1}-z_i^{-\lambda_j-n+j-1})}{\det(z_i^{n-j+1}-z_i^{-n+j-1})}$,
and $\omega=e^{i\pi/3}$;
this formula can be proved by induction, 
or can be seen as a byproduct of this paper,
as we now show. 

In the limit $\zeta\to 0$, the parameterization $w$ is related to the
multiplicative spectral parameter $z$
by $w=(z-1)^2/(1+z+z^2)$; this way we find
\[
h(z,z')=\frac{9(z^2+zz'+z'^2)(1+zz'+z^2z'^2)}{(1+z+z^2)^2(1+z'+z'^2)^2}
\]
The denominator factors out of Pfaffians and determinants.

\subsection{Pfaffians}
We now recognize the Pfaffian $A_n$ (Eq.~\eqref{defA}) in even size:
\begin{multline*}
A_{2m}(w_1,\ldots,w_{2m})=
3^m\prod_{i=1}^n z_i 
\prod_{1\le i<j\le 2m} \frac{3(z_i^2+z_iz_j+z_j^2)(1+z_iz_j+z_i^2z_j^2)}{(1+z_i+z_i^2)(1+z_j+z_j^2)(z_i-z_j)(1-z_iz_j)}
\\
\Pf \frac{(z_i-z_j)(1-z_iz_j)}{(z_i^2+z_iz_j+z_j^2)(1+z_iz_j+z_i^2z_j^2)}
\end{multline*}
which up to some prefactors is exactly the Pfaffian given in \cite{PDF-open}
(Eq.~(3.27)) for the {\em square}\/ of the partition function $Z_{UASM}$
of U-turn symmetric ASMs of \cite{kup-symASM}.
The latter is known to coincide with $\chi_{Y_{2m}}(z_1,\ldots,z_{2m})$ 
\cite{Oka} and
so we reproduce the first factor of the l.h.s.\ of Eq.~\eqref{halfspectrig}.
More precisely, we find
$
A_{2m}(w_1,\ldots,w_{2m})=
3^{2m^2}\prod_{i=1}^{2m} (1+z_i+z_i^{-1})^{-2m+1}\chi_{Y_{2m}}(z_1,\ldots,z_{2m})^2
$.
The odd case can be reduced to the even case by sending one of the $z_i$
to zero (something which did not make sense in the elliptic setting),
so that for both parities we have
\[
A_n(w_1,\ldots,w_n)=3^{2\lfloor n/2 \rfloor \lfloor(n+1)/2\rfloor}
\prod_{i=1}^{n} (1+z_i+z_i^{-1})^{-n+1}
\chi_{Y_{n}}(z_1,\ldots,z_{n})^2
\]
or in terms of the original quantities,
$\A_n(x_1,\ldots,x_n)=3^{-2\lfloor n/2\rfloor \lfloor (n-1)/2\rfloor}\chi_{Y_n}(z_1,\ldots,z_n)^2$.

The second factor is simply obtained by noting that
$w=J_2=-1/2$ corresponds to $z=\omega=e^{i\pi/3}$,
so 
\[
B_n(w_1,\ldots,w_n)=2^{-n}3^{2\lfloor n/2+1 \rfloor \lfloor(n+1)/2\rfloor}
\prod_{i=1}^{n} (1+z_i+z_i^{-1})^{-n}
\chi_{Y_{n+1}}(z_1,\ldots,z_n,\omega)^2
\]
or $\B_n(x_1,\ldots,x_n)=3^{-2\lfloor n/2 \rfloor \lfloor(n+1)/2\rfloor}
\chi_{Y_{n+1}}(z_1,\ldots,z_n,\omega)^2$.
Finally, 
\[
\Z_n=\prod_{i=1}^n \left(\frac{1+z_i+z_i^{-1}}{3}\right)^2\X_n=
3^{-n^2}
\prod_{i=1}^n (1+z_i+z_i^{-1})^2 \chi_{Y_n}(z_1,\ldots,z_n)^2
\chi_{Y_{n+1}}(z_1,\ldots,z_n,\omega)^2
\]
which is consistent with Eqs.~\eqref{Ztrig} and \eqref{halfspectrig}.

\subsection{Determinants}
Similarly, the determinants simplify as $\zeta\to 0$. Noting that
$w=J_3$ and $w=J_4$ both correspond to $z=0$, we conclude that there
are only two distinct determinants for each parity;
Tsuchiya's determinant
\cite{Tsuchiya,kup-symASM} is known to be equal at a cubic root of unity to the symplectic character introduced above \cite{Oka}
\begin{multline*}
\frac{
\prod_{\substack{1\le i\le m,\\m+1\le j\le 2m}}(z_i^2+z_iz_j+z_j^2)(1+z_iz_j+z_i^2z_j^2)}
{\prod_{\substack{1\le i<j\le m,\\m+1\le i<j\le 2m}} (z_j-z_i)(1-z_iz_j)}
\det_{\substack{1\le i\le m,\\m+1\le j\le 2m}}
\frac{1}{(z_i^2+z_iz_j+z_j^2)(1+z_iz_j+z_i^2z_j^2)}
\\
=\chi_{Y_{2m}}(z_1,\ldots,z_{2m})
\end{multline*}
and then we have:
\begin{align*}
H_{2m}(w_1,\ldots,w_{2m})
&=3^{m(m-1)}\prod_{i=1}^{2m} (1+z_i+z_i^{-1})^{-m+1} \chi_{Y_{2m}}(z_1,\ldots,z_{2m})
\\
H_{2m}(w_1,\ldots,w_{2m-1},J_2)
&=3^{m(m-1)}\prod_{i=1}^{2m-1} (1+z_i+z_i^{-1})^{-m+1} \chi_{Y_{2m}}(z_1,\ldots,z_{2m-1},\omega)
\\
H_{2m}(w_1,\ldots,w_{2m-1},J_{3/4})
&=3^{m(m-1)}\prod_{i=1}^{2m-1} (1+z_i+z_i^{-1})^{-m+1} \chi_{Y_{2m-1}}(z_1,\ldots,z_{2m-1})
\\
H_{2m+2}(w_1,\ldots,w_{2m},J_2,J_{3/4})
&=3^{m(m+1)}2^{-m}\prod_{i=1}^{2m} (1+z_i+z_i^{-1})^{-m} \chi_{Y_{2m+1}}(z_1,\ldots,z_{2m},\omega)
\\
H_{2m+2}(w_1,\ldots,w_{2m},J_3,J_4)
&=3^{m(m+1)}\prod_{i=1}^{2m} (1+z_i+z_i^{-1})^{-m} \chi_{Y_{2m}}(z_1,\ldots,z_{2m})
\\
H_{2m+2}(w_1,\ldots,w_{2m-1},J_2,J_3,J_4)
&=3^{m(m+1)}\prod_{i=1}^{2m-1} (1+z_i+z_i^{-1})^{-m} \chi_{Y_{2m}}(z_1,\ldots,z_{2m-1},\omega)
\end{align*}

\subsection{More determinants}
The expression \eqref{slater} of $\S_{2m}$ as a Slater determinant reduces to the
numerator of our definition of the symplectic character $\chi_{Y_{2m}}$ (since $k_j=Y_{2m;m+1-j}+2m-j+1$, $j=1,\ldots,2m$)
\[
\S_{2m}(z_1,\ldots,z_{2m})=\det_{i,j=1,\ldots,2m} (z_i^{k_{j}}-z_i^{-k_{j}})
\]
The differential equation \eqref{prediffeqa} reduces to
\[
\sum_{i=1}^{2m} \left(z_i\frac{\der}{\der z_i}\right)^2 \S_{2m}(z_1,\ldots,z_{2m})
=m(6m^2-1)\S_{2m}(z_1,\ldots,z_{2m})
\]

\subsection{Homogeneous limit}
Finally, 
$A_{2m}^{1/2}=H_{2m}=3^{-m(m-1)} \chi_{Y_{2m}}(1,\ldots,1)=1,1,3,26,646\ldots$ is the number of Vertically Symmetric Alternating Sign Matrices of size $2m+1$ (also, the number of 
Off-diagonally Symmetric Alternating Sign Matrices of size $2m$,
and the number of Descending Plane Partitions of size $m$ which are
symmetric w.r.t.\ all reflections, i.e.,
Cyclically Symmetric Transpose Complement Plane Partitions of a hexagon of size $(m+1)\times(m-1)$ with a triangular hole cut out),
while $A_{2m-1}^{1/2}=H_{2m}(J_{3/4})=3^{-(m-1)^2} \chi_{Y_{2m-1}}(1,\ldots,1)=1,2,11,170\ldots$ is the number
of Cyclically Symmetric Transpose Complement Plane Partitions of size $m$
(also, the number of VSASMs of size $(2m-1)\times(2m+1)$ 
with a defect on the $m^{\mathrm{th}}$ row, the symmetry line).
Note that the square of the number of VSASMs also appears in the observations
of \cite{RS-XYZ}.

The sequence of numbers $2^{m}B_{2m}^{1/2}=2^{m}H_{2m}(J_2,J_{3/4})=3^{-m(m-1)}\chi_{Y_{2m+1}}(1,\ldots,1,\omega)=1,5,66,2431\ldots$ appears as one of the factors of the enumeration
of UUASMs in \cite{kup-symASM}. The last sequence, $2^m (B_{2m-1}/3)^{1/2}=H_{2m}(J_2)=3^{-(m-1)^2}\chi_{Y_{2m}}(1,\ldots,1,\omega)=1,7,143,8398,\ldots$ is the number of ASMs of order $2m+1$ divided by the number of VSASMs of size $2m+1$.

As mentioned before, the last two cases, namely $H_{2m}(J_3,J_4)$, and
$H_{2m}(J_2,J_3,J_4)$, are related to $H_{2m}$ and $H_{2m}(J_2)$ by multiplication
by powers of $3$ and $2$.

\section{The $\zeta\to 1$ limit}\label{trivtrigo}
Besides the $\zeta\to0$ limit, there is another trigonometric limit,
namely
$\zeta\to1$ or $\alpha\to0$. It is expected to be somewhat trivial since
the correponding Hamiltonian is the Ising Hamiltonian with interaction
$\sigma^x\sigma^x$. Indeed, we find that the building block $H_{2m}$
of the partition function becomes:
\[
H_{2m}(w_1,\ldots,w_{2m})|_{\zeta=1}=
\frac{\displaystyle
\prod_{\substack{1\le i\le m\\m+1\le j \le 2m}}(1-4 w_i w_j)}
{\displaystyle
\prod_{\substack{1\le i<j\le m\\\text{or}\\m+1\le i<j\le 2m}}(w_i-w_j)}
\det_{\substack{1\le i\le m\\m+1\le j\le 2m}} \frac{1}{1-4 w_i w_j}
=2^{m(m-1)}
\]
This formula is valid as long as the $w_i$ stay finite as $\zeta\to1$. One special case is if one $w_i$ is equal to
$J_4=1/(1-\zeta)$. Then we find instead
\[
H_{2m}(w_1,\ldots,w_{2m-1},J_4)|_{\zeta=1}=2^{(m-1)^2}
\]

so that
\[
X_n(\ldots)=2^{n(n+1)+1}
\]
This is compatible with a constant value of $\Psi_{n,\alpha}=2^{n(n-1)/2}$ since $X_{2m}=2^{2n+1} \Psi_{2m,\alpha}^2$.

\vfill\eject
\section{Proof of symmetry of $H_{2m}$}\label{proofsym}
The symmetry of $\H_{2m}$, defined by \eqref{detH} can be seen
as a particular case of a general result, which can be formulated
as follows: (see also Thm.~4.2 in \cite{Oka-minor})
\begin{proposition*}
Let $\phi_1,\phi_2$ be two functions (with values in $\mathbb{C}$) 
such that\break
(i) $\phi(x,y)=-\phi(y,x)$, (ii) $\phi(x_1,x_2)\phi(x_3,x_4)-\phi(x_1,x_3)\phi(x_2,x_4)+\phi(x_1,x_4)\phi(x_2,x_3)=0$ for $\phi=\phi_1,\phi_2$. Then,
in the domain of the $(x_i)_{1\le i\le 2m}$ such that $\phi_2(x_i,x_j)\ne 0$ for all $1\le i,j\le 2m$,
\[
\Delta_{2m}(x_1,\ldots,x_{2m})=\frac{\displaystyle\det_{\substack{i=1,\ldots,m\\j=m+1,\ldots,2m}}
\left(\frac{\phi_1(x_i,x_j)}{\phi_2(x_i,x_j)}\right)}{\displaystyle\prod_{\substack{1\le i<j\le m\\\mathrm{or}\\m+1\le i<j\le 2m}} \phi_2(x_i,x_j)}
\]
is symmetric in all arguments $\{x_1,\ldots,x_{2m}\}$.
\end{proposition*}
Actually it is well-known that functions that satisfy \textit{(i)} and \textit{(ii)}
are $2\times 2$ determinants $\left|\begin{smallmatrix} a(x)&a(y)\\ b(x)&b(y)\end{smallmatrix}\right|$, so that, removing symmetric factors, one may without
loss of generality write $\phi_i(x_1,x_2)=\phi_i(x_1)-\phi_i(x_2)$, $i=1,2$.
The proposition then follows from the following representation
(characteristic of Toda chain tau functions):
starting from $\frac{\phi_1(x_i)-\phi_1(x_j)}{\phi_2(x_1)-\phi_2(x_j)}
= \frac{1}{2\pi i}\oint_C \frac{dy}{(y-\phi_2(x_i))(y-\phi_2(x_j))}\phi_1(y)$ where
$C$ is any contour that surrounds once counterclockwise the $\phi_2(x_j)$,
$j=1,\ldots,2m$, and expanding the determinant in $\Delta_{2m}$ we get
\begin{align*}
\Delta_{2m}(x_1,\ldots,x_{2m})&=\frac{1}{m!(2\pi i)^m}
\oint_{C^m} \prod_{i=1}^m dy_i \phi_1(y_i) 
\eqbreak
\frac{\displaystyle\det_{\substack{i=1,\ldots,m\\j=1,\ldots,m}}\left(\frac{1}{y_i-\phi_2(x_j)}\right)}{\prod_{1\le i<j\le m}(\phi_2(x_i)-\phi_2(x_j))}
\frac{\displaystyle\det_{\substack{i=1,\ldots,m\\j=m+1,\ldots,2m}}\left(\frac{1}{y_i-\phi_2(x_j)}\right)}{\prod_{m+1\le i<j\le 2m}(\phi_2(x_i)-\phi_2(x_j))}
\\
&=\frac{1}{m!(2\pi i)^m}
\oint_{C^m} \prod_{i=1}^m dy_i \phi_1(y_i)
\frac{\prod_{1\le i<j\le m}(y_i-y_j)^2}{\prod_{i=1}^m \prod_{j=1}^{2m} (y_i-\phi_2(x_j))}
\end{align*}
which is explicitly symmetric in the $x_i$.

The application to $\H_{2m}$ 
consists in writing $\phi_2(x,y)=\h(x,y)\th(x-y)\th(x+y)$,
$\phi_1(x,y)=\th(x-y)\th(x+y)$ and checking that
they satisfy \textit{(i)} and \textit{(ii)}, so that $\H_{2m}(x_1,\ldots,x_{2m})=
\prod_{1\le i<j\le 2m}\h(x_i,x_j)\Delta_{2m}(x_1,\ldots,x_{2m})$.
It is slightly easier to apply it to $H_{2m}$, i.e., after
the change of variables from $x$ to $w$, since we then have the more
explicit expressions
$\phi_1(w)=w$, $\phi_2(w)=w/(1+(3+\zeta^2)w^2-(1-\zeta^2)w^3)$.

Note that other identities following from integrability of the Toda chain,
for example
the Hankel determinant form
\[
\Delta_{2m}(x_1,\ldots,x_{2m})=\det(s_{i+j})_{i,j=0,\ldots m-1},
\qquad s_k=\sum_{i=1}^{2m} \frac{\phi_2(x_i)^k}{\prod_{j(\ne i)}(\phi_2(x_i)-\phi_2(x_j))} \phi_1(x_i)
\]
They also
provide an alternative derivation of Eq.~\eqref{slater} (``first quantized''
form of the tau function).

\section{Differential equations}\label{moremesses}
We provide here analogues of Eqs.~\eqref{bigmess} and \eqref{diffeq} when $H_{2m}$
(that is, the function $H_{2m}$ with all arguments set to zero)
is replaced with $H_{2m}(S)$, $S\subset \{J_2,J_3,J_4\}$
(again, with all other arguments set to zero).
Because of the permutation symmetry w.r.t.\ $\{J_2,J_3,J_4\}$, 
we only need to provide one formula for each possible
cardinality of $S$. When taking derivatives w.r.t.\ $u$ or $v$,
the convention is that the arguments that are specialized to $J_2,J_3,J_4$
are among the $u$'s.

After transposition (for display purposes), Eq.~\eqref{bigmess} is of the form
\[
\left(
\frac{\der^2}{\der u\der v} H_{2m},
\ 
\frac{\der^2}{\der u^2} H_{2m},
\ 
\frac{\der^2}{\der \alpha^2} H_{2m},
\ 
\frac{\der^2}{\der u\der \alpha} H_{2m},
\ 
\frac{\der}{\der u} H_{2m},
\ 
\frac{\der}{\der \alpha} H_{2m},
\ 
H_{2m}
\right)
P=0
\]

For $H_{2m}$ itself, the matrix $P$ is
\[\left(\vcenter{
\halign{&\hfil$\,\scriptstyle #\,$\hfil\crcr
0 & 0 & 2 (2 m+1) & 2 \left(4 m^2+m-2\right) \alpha  \cr
 0 & 0 & 4 m & -2 m (4 m+1) \alpha  \cr
 0 & -2 m (\alpha -1) (\alpha +8) & 0 & 0 \cr
 0 & -2 (-4 m-1) & -2 m (\alpha -1) (\alpha +8) & 0 \cr
 2 (4 m+1) & 0 & m \left(m^2-m+1\right) (\alpha +2) & m (2 m+1)
(\alpha -4)^2 \cr
 -2 m (\alpha -1) (\alpha +8) & m \left(\alpha  m^2+2 m^2-\alpha  m-2
m-4 \alpha -14\right) & 0 & 0 \cr
 (m-1) m^2 (\alpha +2) & (m-1) m^2 & -(m-1) m^2 (\alpha -4) & -(m-1)
m^2 (2 m+1) (\alpha -4) \alpha  \cr
}}\right)\]

For $H_{2m}(J_2)$:
\[\left(\vcenter{
\halign{&\hfil$\,\scriptstyle #\,$\hfil\crcr
0 & 0 & 2 (2 m+1) & 2 \left(4 m^2-5 m-1\right) \alpha  \cr
 0 & 0 & 4 m & -2 m (4 m-1) \alpha  \cr
 0 & -2 (m-1) (\alpha -1) (\alpha +8) & 0 & 0 \cr
 0 & -2 (1-4 m) & -2 m (\alpha -1) (\alpha +8) & 0 \cr
 2(4 m-1) & 0 & m \left(\alpha  m^2+2 m^2-\alpha  m-4 m+\alpha +4\right) & m \left(2 m \alpha ^2-\alpha ^2-16 m \alpha +32 m \right) \cr
 -2 (m-1) (\alpha -1) (\alpha +8) & (m-1) \left(\alpha  m^2+2 m^2-\alpha  m-4 m-4 \alpha -12\right) & 0 & 0 \cr
 (m-1)^2 (\alpha  m+2 m-2) & (m-1)^2 m & -(m-1) m (\alpha  m-4 m+4) &
-(m-1) m^2 \alpha  (2 \alpha  m-8 m-\alpha +8) \cr
}}\right)\]

For $H_{2m}(J_3,J_4)$:
\[\left(\vcenter{
\halign{&\hfil$\,\scriptstyle #\,$\hfil\crcr
0 & 0 & 2 (2 m+1) & 2 \left(4 m^2-11 m+4\right) \alpha  \cr
 0 & 0 & 4 m & -2 m (4 m-3) \alpha  \cr
 0 & -2 (m-2) (\alpha -1) (\alpha +8) & 0 & 0 \cr
 0 & 2 (4 m-3) & -2 m (\alpha -1) (\alpha +8) & 0 \cr
 2 (4 m-3) & 0 & m \left(\alpha  m^2+2 m^2-\alpha  m+\alpha +2\right) 
& m \left(2 m \alpha ^2-\alpha ^2-16 m \alpha +16 \alpha +32 
m-32\right) \cr
 -2 (m-2) (\alpha -1) (\alpha +8) & (m-2) \left(\alpha  m^2+2 
m^2-\alpha  m-4 \alpha -14\right) & 0 & 0 \cr
 (m-2) m (\alpha  m+2 m-\alpha ) & (m-2) (m-1) m & -(m-2) m 
(\alpha  m-4 m-\alpha ) & -(m-2) (m-1) m \alpha  (2 \alpha  m-8 
m-\alpha ) \cr
}}\right)\]

For $H_{2m}(J_2,J_3,J_4)$:
\[\left(\vcenter{
\halign{&\hfil$\,\scriptstyle #\,$\hfil\crcr
 0 & 0 & 2 (2 m+1) & 2 (m-1) (4 m-13) \alpha  \cr
 0 & 0 & 4 m & -2 m (4 m-5) \alpha  \cr
 0 & -2 (m-3) (\alpha -1) (\alpha +8) & 0 & 0 \cr
 0 & 2 (4 m-5) & -2 m (\alpha -1) (\alpha +8) & 0 \cr
 2 (4 m-5) & 0 & m \left(m^2-m+1\right) (\alpha +2) & m (2 m-3) 
(\alpha -4)^2 \cr
 -2 (m-3) (\alpha -1) (\alpha +8) & (m-3) \left(\alpha  m^2+2 
m^2-\alpha  m-2 m-4 \alpha -14\right) & 0 & 0 \cr
 (m-3) (m-1) m (\alpha +2) & (m-3) (m-1) m & -(m-3) m^2 (\alpha -4) & 
-(m-3) m^2 (2 m-3) (\alpha -4) \alpha  \cr
}}\right)\]

As to Eq.~\eqref{diffeq}:
\[
C_0\, H_{2(m+1)}H_{2(m-1)}=
C_1\,H_{2m} H''_{2m} -C_2\, (H'_{2m})^2
+C_3\,H_{2m} H'_{2m}+C_4\, H_{2m}^2 
\]
The coefficients for $H_{2m}$ are:
\begin{align*}
C_0
&=4 \alpha  (4 m-1) (4 m+1)^2 (4
   m+3)
\\
C_1
&=2 (\alpha -1)^2 \alpha  (\alpha +8)^2 (4 m+1)^2\\
C_2
&=4 (\alpha -1)^2 \alpha  (\alpha +8)^2 (4 m-1) (4 m+3)\\
C_3
&= (\alpha -1) (\alpha +8)
   (\alpha ^2+28 \alpha +24 \alpha ^2 m^2+304 \alpha  m^2-256 m^2
+20 \alpha ^2 m+208 \alpha  m-192 m-32)\\
C_4
&=6 \alpha ^2-24
   \alpha +4 \alpha ^3 m^4-1008 \alpha ^2 m^4+3408 \alpha  m^4+512 m^4-4 \alpha ^3 m^3-984 \alpha ^2 m^3
+4032 \alpha  m^3
\eqbreak
-128 m^3
+\alpha
   ^3 m^2-142 \alpha ^2 m^2+1028 \alpha  m^2-320 m^2-\alpha ^3 m+28 \alpha ^2 m-44 \alpha  m-64 m\\
 \end{align*}
For $H_{2m}(J_2)$:
\begin{align*}
C_0
&=4\alpha (4m-3) (4m-1)^2 (4 m+1) 
\\
C_1
&=
2 (4m-1)^2 (-1 + \alpha)^2 \alpha (8 + \alpha)^2
\\
C_2
&=
4 (4m-3) (1 + 4 m) (-1 + \alpha)^2 \alpha (8 + \alpha)^2
\\
C_3
&=
 (-1 + \alpha) (8 + \alpha) (64 m - 256 m^2 - 
   4 \alpha - 96 m \alpha + 304 m^2 \alpha + \alpha^2 - 
   4 m \alpha^2 + 24 m^2 \alpha^2)
\\
C_4
&=
-128 m + 768 m^2 - 1152 m^3 + 512 m^4 - 36 \alpha + 
   336 m \alpha - 204 m^2 \alpha - 2784 m^3 \alpha +
   3408 m^4 \alpha \eqbreak + 18 \alpha^2 - 126 m \alpha^2 - 
   6 m^2 \alpha^2 + 1032 m^3 \alpha^2 - 1008 m^4 \alpha^2 
   -m \alpha^3 + 9 m^2 \alpha^3 - 12 m^3 \alpha^3 + 
   4 m^4 \alpha^3
\\
\end{align*}
For $H_{2m}(J_3,J_4)$:
\begin{align*}
C_0
&=4\alpha (4 m-5) (4m-3)^2 (4m-1) 
\\
C_1
&=
2 (4m-3)^2 (-1 + \alpha)^2 \alpha (8 + \alpha)^2
\\
C_2
&=
4 (4m-5) (4m-1) (-1 + \alpha)^2 \alpha (8 + \alpha)^2
\\
C_3
&=
(-1 + \alpha) (8 + \alpha) (-192 + 448 m - 256 m^2 + 
   204 \alpha - 512 m \alpha + 304 m^2 \alpha + 21 \alpha^2 
\eqbreak
-    44 m \alpha^2 + 24 m^2 \alpha^2)
\\
C_4
&=
384 m^2 - 896 m^3 + 512 m^4 + 720 \alpha - 5208 m \alpha + 
 11892 m^2 \alpha - 10848 m^3 \alpha + 3408 m^4 \alpha \eqbreak- 
 90 \alpha^2 + 1074 m \alpha^2 - 2958 m^2 \alpha^2 + 
 3000 m^3 \alpha^2 - 1008 m^4 \alpha^2 + 3 m \alpha^3 - 
 3 m^2 \alpha^3 - 4 m^3 \alpha^3 + 4 m^4 \alpha^3
\\
\end{align*}
For $H_{2m}(J_2,J_3,J_4)$:
\begin{align*}
C_0
&=4 \alpha  (4 m-7) (4 m-5)^2 (4 m-3)
\\
C_1
&=
2 (4m-5)^2 (-1 + \alpha)^2 \alpha (8 + \alpha)^2
\\
C_2
&=
-4 (4m-7) ( 
   4 m-3) (-1 + \alpha)^2 \alpha (8 + \alpha)^2
\\
C_3
&=
 (-1 + \alpha) (8 + \alpha) (-480 + 704 m - 256 m^2 + 
   540 \alpha - 816 m \alpha + 304 m^2 \alpha
\eqbreak
 + 45 \alpha^2 - 
   68 m \alpha^2 + 24 m^2 \alpha^2)
\\
C_4
&=
 -960 m + 2368 m^2 - 1920 m^3 + 512 m^4 + 8400 \alpha - 
   27740 m \alpha + 33572 m^2 \alpha 
\eqbreak
- 17664 m^3 \alpha + 
   3408 m^4 \alpha - 2100 \alpha^2 + 7240 m \alpha^2 - 
   9142 m^2 \alpha^2 
+ 5016 m^3 \alpha^2 \eqbreak
- 1008 m^4 \alpha^2 - 
   5 m \alpha^3 + 13 m^2 \alpha^3 - 12 m^3 \alpha^3 + 
   4 m^4 \alpha^3
\\
\end{align*}

\def\urlshorten http://#1/#2urlend{{#1}}%
\renewcommand\url[1]{%
\href{#1}{\scriptsize\expandafter\path\urlshorten#1 urlend}%
}
\gdef\MRshorten#1 #2MRend{#1}%
\gdef\MRfirsttwo#1#2{\if#1M%
MR\else MR#1#2\fi}
\def\MRfix#1{\MRshorten\MRfirsttwo#1 MRend}
\renewcommand\MR[1]{\relax\ifhmode\unskip\spacefactor3000 \space\fi
\MRhref{\MRfix{#1}}{{\scriptsize \MRfix{#1}}}}
\renewcommand{\MRhref}[2]{%
\href{http://www.ams.org/mathscinet-getitem?mr=#1}{#2}}

\bibliography{biblio}
\bibliographystyle{amsplainhyper}

\end{document}